 \documentclass[12pt]{article}

 \usepackage{amssymb}
 \newcommand{\E}{\mathbb{E}}

\newcommand{\R}{\mathbb{R}}
\newcommand{\Z}{\mathbb{Z}}

\newcommand{\C}{\mathbb{C}}

 \newcommand{\beqn}{\begin{eqnarray}}
 \newcommand{\eeqn}{\end{eqnarray}}
 \newcommand{\be}{\begin{equation}}
 \newcommand{\ee}{\end{equation}}
 \newcommand{\ba}{\begin{array}}
 \newcommand{\ea}{\end{array}}

  \newcommand{\ov}{\overline}
 \newcommand{\ve}{\varepsilon}
 \newcommand{\ds}{\displaystyle}
 \newcommand{\ti}{\tilde}
 \newcommand{\na}{\nabla}
 
 \newcommand{\br}{|\kern-.25em|\kern-.25em|}
 \newcommand{\brr}{{|\kern-.15em|\kern-.15em|\kern-.15em}\,}

\newcommand{\mod}{\mathop{\rm mod}\nolimits}
\newcommand{\supp}{\mathop{\rm supp}\nolimits}
\newcommand{\sgn}{\mathop{\rm sgn}\nolimits}

 \newtheorem{theorem}{Theorem}[section]
 \newtheorem{definition}[theorem]{Definition}
 
 \newtheorem{lemma}[theorem]{Lemma}
 
 \newtheorem{remark}[theorem]{Remark}
 
 \newtheorem{cor}[theorem]{Corollary}
 \newtheorem{pro}[theorem]{Proposition}

\newcommand{\bo}{{\hfill\loota}}
\newcommand{\loota}{\hbox{\enspace{\vrule height 7pt depth 0pt width
      7pt}}}

\begin{document}

\begin{titlepage}
\hspace{2cm}
 \begin{center}
{\Large\bf On the convergence to a statistical equilibrium
\medskip\\
  for the wave equations coupled to a particle}
\vspace{1cm}\\
{\large T.V. Dudnikova}
 \footnote{Electronic mail:~tdudnikov@mail.ru}\medskip\\
{\it  Elektrostal Polytechnical Institute,
Elektrostal 144000, Russia}
\end{center}
\vspace{1cm}

\begin{abstract}
We consider a linear Hamiltonian system consisting of
a classical particle and a scalar field describing
by the wave or Klein-Gordon equations with variable coefficients.
The initial data of the system are supposed to be a
random function which has  some mixing properties.
  We study the distribution $\mu_t$
  of the random solution at time moments $t\in\R$.
  The main result is the convergence of $\mu_t$
   to a Gaussian probability measure as $t\to\infty$.
The mixing properties of the limit measures are studied.
The application to the case of Gibbs initial measures is given.
\medskip

\medskip

\end{abstract}
\end{titlepage}

%%---------------------------------------------
\section{ Introduction}
%%-------------------------------------------
The paper concerns  problems of long-time convergence to an
 equilibrium distribution for a coupled system consisting
of a field and a particle. For  one-dimensional chains of harmonic
oscillators, the results have  been established by Spohn and
Lebowitz in \cite{SL}, and by Boldrighini {\it et al.} in \cite{BPT}.
Ergodic properties of one-dimensional chains of anharmonic
oscillators coupled to heat baths were studied by Jak\v{s}i\'c, Pillet
and others (see, e.g., \cite{JP98,EPR}). In
\cite{DKKS,DKRS,DKS,DKM2}, we studied the convergence to
equilibrium for the systems described by partial differential
equations. Later on, similar results were obtained in
\cite{DKS1} for $d$-dimensional harmonic crystals with $d\ge1$,
  and in \cite{DK} for a scalar field coupled to a harmonic crystal.

 Here we treat the linear
Hamiltonian system consisting of the scalar wave or Klein--Gordon
field $\varphi(x)$, $x\in\R^d$, coupled to a classical particle
with position in $q\in\R^d$, $d\ge3$. The Hamiltonian functional
of the coupled system reads
 \be\label{Ham}
 H(\varphi,\pi,q,p)=H_A(q,p)+H_B(\varphi,\pi)
 +q\cdot\langle \nabla\varphi,\rho\rangle.
 \ee
 Here "$\cdot$" stands for the standard Euclidean scalar product in
 $\R^d$, $\langle\cdot,\cdot\rangle$ denotes the inner product
 in the real Hilbert space $L^2(\R^d)$ (or its extensions),
  $H_A$ is the Hamiltonian of the particle,
$$
  H_A(q,p)=\frac12 \Big(|p|^2+\omega^2 |q|^2\Big),
  \quad \mbox{with some }\,\,\omega>0,
 $$
and $H_B$ denotes the Hamiltonian for the wave or Klein-Gordon field.
We suppose that
$$
H_B(\varphi,\pi)=\frac12 \int\limits_{\R^d}
 \Bigr( \sum\limits_{i,j=1}^d a_{ij}(x)
 \nabla_i\varphi(x)\nabla_j \varphi(x)
 +a_0(x)|\varphi(x)|^2+|\pi(x)|^2\Bigr)\,dx
$$
 in the case of the wave field (WF), and
$$
H_B(\varphi,\pi)= \frac12 \int\limits_{\R^d}
 \Bigr( \sum\limits_{j=1}^d|(\nabla_j-iA_j(x))
  \varphi(x)|^2+ m^2|\varphi(x)|^2+|\pi(x)|^2\Bigr)\,dx,
  \quad \mbox{with some }\,\,m>0,
 $$
 in the case of the Klein-Gordon field (KGF).
 We impose the conditions {\bf A1}--{\bf A5} below on the coefficients
 $a_{ij}(x)$, $a_0(x)$
 and $A_j(x)$. In particular, the functions $a_{ij}(x)-\delta_{ij}$, $a_0(x)$
 and $A_j(x)$ vanish outside a bounded domain.
 %In the first case, we assume that
 %$a_{ij}(x),a_0(x)\in C^\infty(\R^d)$; $a_{ij}(x)=\delta_{ij}$
 %and $a_0(x)=0$ for $|x|>R_a$.
%In the second case,  $m>0$ is a fixed constant,
 %the functions $A_j(x)\in C^\infty(\R^d)$ and
 %$A_j(x)=0$ for $|x|>R_a$.
 % for $d=2$, $\nabla_2 A_1\not\equiv\nabla_1 A_2$.
 %Moreover,  $d\ge3$, and $d$ is odd in the case of the WF.
%%-------------------------
\medskip

     We assume that the initial data
  $Y_0:=(\varphi_0,\pi_0,q_0,p_0)$ are a
random element of a real functional space ${\cal E}$ consisting of
states with finite local energy, see Definition \ref{d1.1'}
below. The distribution of $Y_0$ is a probability measure  $\mu_0$
of mean zero satisfying conditions {\bf S1}--{\bf S3}  below. In
particular, we assume that
 the initial measure $\mu_0$ satisfies a mixing condition.
Roughly speaking, it means that
$$
  Y_0(x)\quad \mbox{and }\,\,\, Y_0(y)
\quad  \mbox{are asymptotically  independent as }\,\,
|x-y|\to\infty.
$$

 We study the distributions $\mu_t$, $t\in\R$, of the random
solution $Y_t:=(\varphi_t,\pi_t,q_t,p_t)$ at time moments
$t\in\R$. Our main objective is  to prove the weak convergence of
the measures $\mu_t$  to an equilibrium measure $\mu_{\infty}$,
\begin{equation}\label{1.8i}
  \mu_t \rightharpoondown \mu_\infty\quad  \mbox{as }\,\,t\to \infty,
  \end{equation}
  where the limit measure $\mu_{\infty}$ is
 a Gaussian measure on ${\cal E}$.
 We derive the explicit formulas for the limiting
 correlation functions of
$\mu_\infty$. The similar convergence holds for  $t\to-\infty$
because our system is time-reversible.
We prove that the dynamic group is mixing (and, in particular, ergodic)
with respect to the limit measures $\mu_\infty$.
Moreover, we extend results to the case of
non translation-invariant initial measures $\mu_0$
and give an application to the case of the Gibbs initial measures.
\medskip

Let us outline the strategy of the proof.
 When the field variables $(\varphi_t,\pi_t)$
are eliminated from the equations of the coupled system, the
particle evolves according to a linear Volterra
integro-differential equation of a form (see Eqn (\ref{cs2}) below)
  \begin{equation}\label{Lang}
  \ddot q_t=-\omega^2 q_t+
   \int\limits_{0}^t D(t-s)q_s\,ds+F(t),\quad t\in\R,
  \end{equation}
where $D(t)$ is a matrix-valued function depending
on the coupled function $\rho$,
$F(t)$ is a vector-valued function depending on the initial field data
$(\varphi_0,\pi_0)$. Therefore, our first objective is to study
the long-time behavior of the solutions to Eqn (\ref{Lang}). We prove
that for the solutions $q_t$ of Eqn (\ref{Lang}) with $F(t)\equiv0$,
the following bound holds
\be\label{0.2}
|q_t|+|\dot q_t|\le C \ve_F(t),
\ee
 where $\ve_F(t)=e^{-\delta|t|}$ with some $\delta>0$ for the WF,
 and $\ve_F(t)=(1+|t|)^{-3/2}$ for the KGF (see Theorem~\ref{h-eq} below).

 The {\it deterministic} dynamics of the equations with delay
 has been extensively studied by many authors
 under some restrictions on the kernel $D(t)$:
Myshkis \cite{Myshkis}, %(1972)
 Grossman and Millet~\cite{GrossMillet},  %(1973)
   Driver \cite{Driver} %(1977)
  and others.
 For details on the first results and problems
 in the theory of equations with delay, we refer  to
 the survey paper by Corduneanu and Lakshmikantham~\cite{CL}. %(1980)
 For further development of the theory,
see the monograph by Gripenberg, Londen and Staffans~\cite{GLS}. %(1990)
The stability properties for Volterra integro-differential  equations
can be found in the papers
by Murakami %(1991)
\cite{Murakami}, Hara %(1994)
\cite{Hara}, and Kordonis and Philos %(1999)
\cite{KorPhi}.

The linear {\em stochastic} Volterra equations of convolution type
have been treated  also by many authors, see, e.g.,
 Appleby and Freeman %(2003)
   \cite{AF},  the survey article by Karczewska %(2007)
   \cite{Kar}   and the references therein.

Note that in the literature frequently
 the asymptotic behavior of the solutions of Eqn (\ref{Lang})
 is studied assuming that $F(t)$ is a Gaussian with noise or (and)
that the kernel $D(t)$ has the exponential decay or is of one sign.
  However, in our case, $F(t)$ is not Gaussian white-noise, in general.
  Moreover, in the case of the KGF, the decay of $D(t)$ is like $(1+|t|)^{-3/2}$.
\medskip

In recent years the nonlinear {\em generalized Langevin equation},
i.e., the equation of a form (cf. Eqn (\ref{Langevin}) below)
 \begin{equation}\label{N-Lang}
  \ddot q_t=-\nabla V(q_t)-
   \int\limits_{0}^t \Gamma(t-s) \dot q_s\,ds+F(t),
    \quad t\in\R,
     \end{equation}
 with a stationary Gaussian process $F(t)$ and with a smooth
 (confining or periodic) potential $V(q)$,
  has been investigated
 also extensively, see, e.g., \cite{JP97, OP, Snook, Zw73}. In
 particular, the ergodic properties of (\ref{N-Lang}) were studied
 by Jak\v{s}i\'c and  Pillet in \cite{JP97}, the qualitative properties of
 solutions to Eqn~(\ref{N-Lang}) were established by Ottobre
 and Pavliotis  in \cite{OP}.
Rey-Bellet and Thomas \cite{RT} have investigated a model
consisting of a chain of non-linear oscillators coupled to two heat reservoirs.
The nonlinear stochastic integro-differential equations were studied
also in Mao' works (see, e.g., \cite{Mao2000,MR2006}).
\medskip

In this paper, we study a linear "field-particle" model. However,
we do not assume that the initial distribution of the system is a
Gibbs measure or absolutely continuous with respect to a Gibbs
measure.  % or "close to equilibrium".
Therefore, in particular, the
force $F(t)$ in Eqn~(\ref{Lang}) is non-Gaussian, in general.
\medskip

%%------------------------------
 The key step in our proof is the derivation of  the asymptotic
 behavior for the solutions $Y_t$ of the coupled field-particle system.
 Using bound (\ref{0.2}), we prove
the following asymptotics in mean (see Corollary~\ref{c7.2} below):
\begin{equation}\label{0.1}
  \langle Y_t,Z\rangle\sim \langle
W_t(\varphi_0,\pi_0),\Pi(Z)\rangle, \quad t\to\infty,
 \end{equation}
  where $W_t$ is a solving operator to the Cauchy problem for the
wave or Klein-Gordon equations (\ref{KG-vc}),
$(\varphi_0,\pi_0)$ is a initial state of the field,
 and the function $\Pi(Z)$ is defined in (\ref{hn}).
   This asymptotics allows us to apply the results from
    \cite{DKS,DKM2}, where the weak convergence
of the statistical solutions has been proved for wave and
Klein--Gordon equations with variable coefficients.
We divide the proof of
(\ref{1.8i}) into two steps: we first establish the weak
compactness of the measures family $\{\mu_t,\,t\in\R\}$
 (see Section \ref{sec-com}), and then we prove the
convergence of the characteristic functionals of the measures $\mu_t$
(Section \ref{sec-char}).
\medskip

In conclusion, note that convergence (\ref{1.8i}) remains true for
a linear Hamiltonian system consisting of $N$ wave fields
coupled to a single particle.
In this case, the Hamiltonian is
$$
   \sum\limits_{k=1}^N H_B(\varphi_k,\pi_k)+ H_A(q,p)+
 q\cdot\sum\limits_{k=1}^N\langle
 \nabla\varphi_k,\rho_k\rangle.
  $$

 The paper is organized as follows. In Section~2 we describe the
model, impose the conditions on the coupled function $\rho$ and on
the initial measures $\mu_0$ and state the main results.
The limit behavior for solutions
of Eqn (\ref{Lang}) is studied in Section~\ref{sec3}.
 In Section~\ref{sec-com} we prove the compactness of the measures
family $\{\mu_t,\,t\in\R\}$.  The asymptotics (\ref{0.1}) is proved
 in Section~\ref{sec5}.
In Section~\ref{sec-char} we establish the convergence
of characteristic functionals of $\mu_t$  to a limit and complete
the proof of the main result. In Section~\ref{sec7} we study the mixing
properties of the dynamics with respect to the limit measures $\mu_\infty$.
 In Section \ref{sec8} we extend the results to
the case of non translation--invariant initial measures.
Appendix~A concerns the case of Gibbs initial measures.
The existence of the solutions of the coupled system
is proved in Appendix~B.

\newpage
%%%%%%%%%%%%%%%%%%%%%%%%%
\setcounter{equation}{0}
\section{Main Results}
%%-------------------------------------------
\subsection{Model}
%%------------------------

 After taking formally variational derivatives in (\ref{Ham}),
    the coupled dynamics becomes
    %we derive the following equations of motion,
   \beqn\label{1''}
\left.\ba{ll}
  \dot \varphi_t(x) = \pi_t(x),&
  \dot \pi_t(x) = L_B\,\varphi_t(x)+q_t\cdot\nabla\rho(x),
\quad x\in\R^d,\quad t\in\R,\\ &\\
  \dot q_t = p_t,&
 \dot p_t = -\omega^2q_t+\ds\int\limits_{\R^d}
 \nabla\rho(x) \varphi_t(x)\,dx.
 \ea\right|
\eeqn
%%----------------------------------
Here $L_B$ is a differential operator of one of two types:
\be\label{1.2}
L_B=\left\{\ba{l}
L_W:=\sum\limits_{i,j=1}^d
\nabla_i \left(a_{ij}(x)\nabla_j\right) -a_0(x), \\
L_{KG}:=\sum\limits_{j=1}^{d}(\nabla_j-iA_j(x))^2-m^2,
\ea\right.
\ee
where $\nabla_i=\partial/\partial x_i$, $i=1,\dots,d$;
$d\ge 3$, and $d$ is odd in the case when $L_B=L_W$.
%%%------------------------------------------------------
For simplicity of exposition, we consider the case $d=3$ only.
\medskip

We study the Cauchy problem for the system (\ref{1''}) with initial data
\begin{equation}\label{5''}
\varphi_t(x)|_{t=0}=\varphi_0(x),
 \quad \pi_t(x)|_{t=0}=\pi_0(x),
 \quad x\in\R^3,
  \quad q_t|_{t=0}=q_0,
   \quad p_t|_{t=0}=p_0.
 \end{equation}
Write $\phi_t=(\varphi_t(\cdot),\pi_t(\cdot))$,
$\xi_t=(q_t,p_t)$, $Y_t=(\phi_t,\xi_t)$. Then  the system
(\ref{1''})--(\ref{5''}) becomes
 \begin{equation}\label{1.1'}
 \dot Y_t= {\cal L}(Y_t),\quad t\in\R;\quad Y_t|_{t=0}=Y_0.
 \end{equation}

 We assume that the coefficients of $L_B$ satisfy
 the following conditions {\bf A1}--{\bf A5}.\smallskip\\
{\bf A1}. $a_{ij}(x),a_0(x),A_j(x)$ are real  $C^\infty$-functions.\\
{\bf A2}. $a_{ij}(x) = \delta _{ij}$, $a_0(x)=0$,
$A_j(x)=0$  for $|x|>R_a$, where  $R_a<\infty$. Then
$$
L_B\varphi_t(x) =(\Delta-m^2)\varphi_t(x)\quad
     \mbox{for }\,\,|x|>R_a.
$$
Here $m>0$ in the case of the Klein-Gordon field (KGF), i.e., $L_B=L_{KG}$,
   and $m=0$ in the case of the wave field (WF), i.e.,  $L_B=L_W$.
\medskip

In the WF case,  we impose the next conditions {\bf A3} and {\bf A4}.\\
{\bf A3}. $a_0(x)\geq 0$,
and the hyperbolicity condition holds: %$\exists\alpha>0$
 there exists a constant $\alpha>0$ such that
\be\label{1.3}
 \sum_{i,j=1}^3 a_{ij}(x) k_i  k_j\geq \alpha\, |k|^2,
\quad x,k\in\R^3.
\ee
{\bf A4}. A non-trapping condition \cite{V89}:
for $(x(0), k(0))\in\R^3\times\R^3$ with $ k(0)\neq 0$,
\be\label{1.4}
\vert x(t)\vert
\rightarrow \infty\quad \mbox{as }\,\, t\rightarrow \infty,
\ee
where $(x(t), k(t))$ is a solution to the Hamiltonian system
$$ \dot x(t)=\,\,\na_k h(x(t),k(t)),~~
\dot  k(t)=-\na_x h(x(t), k(t)),\quad \mbox{with }\,
h(x,k)=\frac12
 \sum_{i,j=1}^3 a_{ij}(x) k_i  k_j.
$$
{\bf Example}. In the WF case,
{\bf A1}--{\bf A4} hold for the acoustic equation
with constant coefficients
$$
\ddot \varphi_t(x)=\Delta \varphi_t(x),\quad  x\in \R^3.
 $$
 For instance, {\bf A4} follows because $\dot k(t)\equiv0
\Rightarrow x(t)\equiv k(0)t + x(0)$.
\smallskip

%Put $b_{ij}(x)=a_{ij}(x)-\delta_{ij}$.
Write $M_a=\max\limits_{x\in\R^3}
\max\limits_{i,j}\{|a_{ij}(x)-\delta_{ij}|,|a_0(x)|\}$,
 or $M_a=\max\limits_{x\in\R^3}\max\limits_j|A_j(x)|$.\smallskip\\
{\bf A5}. $M_a$ is sufficiently small
%$\max\limits_{x\in\R^3}|a(x)|\ll1$
(we will specify this condition in the proof of Lemma \ref{detA}).
%see bound (\ref{3.19}) below.
\medskip

 Now we formulate the conditions {\bf R1}--{\bf R3} on $\rho(x)$ and $\omega>0$.\smallskip\\
{\bf R1}. In the case of the WF, we assume that
$\Vert\rho\Vert^2_{L^2}< \alpha\,\omega^2$
with $\alpha$ from condition (\ref{1.3}).
In the KGF case, $\Vert\nabla\rho\Vert^2_{L^2}<  m^2\,\omega^2$.
\smallskip\\
{\bf R2}.   The function $\rho(x)$ is a real-valued smooth function,
$\rho(-x)=\rho(x)$, $\rho(x)=0$ for $|x|\ge R_{\rho}$.\\
 {\bf R3}. For any $k\in\R^3\setminus\{0\}$,
   $ \hat\rho(k)=\int e^{ik\cdot x}\rho(x)\,dx\not=0$.
   \medskip\\
%%%------------------------------------------------------------------
{\bf Remark}. Condition {\bf R1} implies that the Hamiltonian $H(\phi_t,\xi_t)$ is nonnegative for finite energy solutions
(see Appendix~B).
In the case of the constant coefficients, i.e.
$L_B=\Delta-m^2$, condition {\bf R1} can be weakened as follows.\\
{\bf R1'}. The matrix $\omega^2I-K_m$ is positive definite, where
  $K_m=(K_{m,ij})_{i,j=1}^3$ stands for the $3\times 3$
 matrix with matrix elements $K_{m,ij}$,
\begin{equation}\label{Km}
   K_{m,ij}:=(2\pi)^{-3}
 \int_{\R^3}\frac{k_ik_j|\hat\rho(k)|^2}{k^2+m^2}\,dk,\quad m\ge0.
\end{equation}
 However, to prove the main result in the case of the KGF,
 we need a stronger condition than {\bf R1'}. Namely, the matrix $(\omega^2-m^2)I-K_m$ is positive definite. This condition is fulfilled, in particular, if $\Vert\nabla\rho\Vert^2_{L^2}<  m^2(\omega^2-m^2)$.

 %%-----------------------  2.2----------
\subsection{Phase space for the coupled system}
%%------------------------
%Let $\Vert\cdot\Vert$ denote the norm in $L^2(\R^3)$.
We introduce a phase space ${\cal E}$.
%%-----------------------------------------------------------
\begin{definition} \label{d1.1'}
(i) Choose a function $\zeta(x)\in C_0^\infty(\R^3)$ with
$\zeta(0)\ne 0$. Denote by $H^s_{\rm loc}(\R^3)$, $s\in \R$,
 the local Sobolev spaces, i.e.,
the Fr\'echet spaces of distributions $\varphi\in D'(\R^3)$
 with the finite seminorms
 $\Vert \varphi\Vert_{s,R}:=
 \Vert\Lambda^s\left(\zeta(x/R)\varphi\right)\Vert_{L^2(\R^3)}$,
  where $\Lambda^s$ stands for the pseudodifferential operator
  with the symbol $\langle  k\rangle^s$, i.e.,
$$
  \Lambda^s \psi:=F^{-1}_{k\to x}(\langle  k\rangle^s\hat
\psi(k)), \quad \langle  k\rangle:=\sqrt{|k|^2+1},
$$
  and $\hat\psi$ is the Fourier transform of the tempered distribution $\psi$.
\medskip\\
(ii) ${\cal H}\equiv H_{\rm loc}^1(\R^3)\oplus H_{\rm loc}^0(\R^3)$
 is the Fr\'echet space
 of pairs $\phi\equiv(\varphi(x),\pi(x))$
with real valued functions $\varphi(x)$ and $\pi(x)$,
 which is endowed with the local energy seminorms
$$
\Vert \phi\Vert^2_{R}= \int\limits_{|x|<R}
(|\varphi(x)|^2+|\nabla\varphi(x)|^2+|\pi(x)|^2) dx<\infty, \quad
 R>0.
 $$
In the case of the KGF, we assume that $\varphi(x)$ and $\pi(x)$ are complex valued functions.
\smallskip\\
(iii) ${\cal E}\equiv {\cal H}\oplus\R^3\oplus\R^3$ is the
Fr\'echet space of vectors $Y\equiv(\phi(x),q,p)$ with the local
energy seminorms
\begin{equation}\label{2.1'}
  \Vert Y\Vert^2_{{\cal E},R}=
\Vert\phi\Vert^2_{R}+|q|^2+|p|^2, \quad  R>0.
  \end{equation}
(iv) For $s\in\R$,
  write ${\cal H}^{s}\equiv H_{\rm loc}^{1+s}(\R^3)
\oplus H_{\rm loc}^{s}(\R^3)$ and
  ${\cal E}^{s}\equiv {\cal H}^{s}\oplus\R^3\oplus\R^3$.
In particular, ${\cal H}\equiv{\cal H}^0$,
 ${\cal E}\equiv{\cal E}^0$.
  \end{definition}
%%-----------------------------------------------------------

  Using the standard technique of pseudodifferential operators and
Sobolev's Theorem (see, e.g., \cite{H3}), one can prove
that
 ${\cal E}^0\equiv{\cal E}\subset {\cal E}^{-\ve}$
  for every $\ve>0$, and the embedding  is compact.
%%--------------------------------------------------------------
\begin{pro}    \label{p1.1'}
Let conditions {\bf A1}--{\bf A3}, {\bf R1} and {\bf R2} hold. Then\\
(i) for every $Y_0 \in {\cal E}$, the Cauchy problem (\ref{1.1'})
has a unique solution $Y_t\in C(\R, {\cal E})$.\\
 (ii) For any $t\in \R$, the operator $S_t:Y_0\mapsto  Y_t$
 is continuous on ${\cal E}$.
 Moreover, for any $T>0$ and $R>\max\{R_\rho,R_a\}$,
$$
   \sup\limits_{|t|\le T}\Vert S_t Y_0\Vert_{{\cal E},R}
 \le C(T)\Vert Y_0\Vert_{{\cal E},R+T}.
$$
\end{pro}

This proposition can be proved using a similar technique
as in \cite[Lemma 6.3]{KSK} and \cite[Proposition 2.3]{D10},
and the proof is based on Lemma \ref{l3.1} below (cf. \cite[Lemma 3.1]{D10}).
 Introduce a Hilbert space $H^1_{F}(\R^3)$ as follows.
 For the KGF,  $H^1_{F}(\R^3)$ is the Sobolev space $H^1(\R^3)$.
In the case of the WF, $H_{F}^1(\R^3)$ stands for
the completion of real space $C_0^\infty(\R^3)$
with norm $\Vert\nabla\varphi\Vert_{L^2}$.
%Then, by Sobolev' embedding theorem,
%$H_F^1(\R^3)=\{\varphi\in L^6(\R^3):\,|\varphi(x)|\in L^2\}$.
Denote by $E$  the Hilbert space
$H^1_{F}(\R^3)\oplus L^2(\R^3)\oplus\R^3\oplus\R^3$
   with finite norm
$$
 \Vert Y\Vert_E^2=\int_{\R^3}\left(|\nabla \varphi(x)|^2+m^2|\varphi(x)|^2
+|\pi(x)|^2\right)dx+|q|^2+|p|^2 \quad \mbox{for }\,
Y=(\varphi(x),\pi(x),q,p),
$$
where $m>0$ for the KGF case, and $m=0$ for the WF case.
%%%--------------------------------------------
\begin{lemma}\label{l3.1}
Let conditions {\bf A1}--{\bf A3}, {\bf R1} and {\bf R2} be valid. Then
the following assertions hold.\\
(i) For every $Y_0\in E$, the Cauchy problem (\ref{1.1'})
has a unique solution $Y_t\in C(\R,E)$. \\
(ii) For $Y_0\in E$,
the energy is conserved, finite and nonnegative,
$H(Y_t)=H(Y_0)\ge0$, $t\in\R$.\\
(iii) For every $t\in \R$, the operator $S_t:Y_0\mapsto Y_t$
is continuous on $E$. Moreover, %for $Y_0\in E$,
\be\label{2.7}
\Vert Y_t\Vert_E\le C\Vert Y_0\Vert_E\quad \mbox{for } \,\,t\in\R.
\ee
%with $B$ depending only on the norm $\Vert Y_0\Vert_E$.
 \end{lemma}

We outline the proof of Lemma \ref{l3.1} and Proposition \ref{p1.1'}
in Appendix B.
%%%------------------------------------------------------------

%%%%%%%%%%%%%%%%%%%%%%%%%%%   2.3    %%%%%%%%%%
\subsection{Conditions on the initial measure}\label{sec2.3}
%%%%%%%%%%%%%%%%%%%%%%%%%%%%%%%%%%%%%%

Let $(\Omega,\Sigma,P)$ be a probability space
 with expectation $\E$
and ${\cal B}({\cal E})$ denote the Borel $\sigma$-algebra in ${\cal
E}$. We assume that $Y_0=Y_0(\omega,x)$ in (\ref{1.1'}) is a
measurable random function with values in
$({\cal E},\,{\cal B}({\cal E}))$.
In other words, $(\omega,x)\mapsto Y_0(\omega,x)$ is a
measurable map $\Omega\times\R^3\to\R^{8}$
 with respect to the (completed) $\sigma$-algebra
$\Sigma\times{\cal B}(\R^3)$ and ${\cal B}(\R^{8})$.
 Then $Y_t=S_t Y_0$ is also a measurable random function with values in
 $({\cal E},{\cal B}({\cal E}))$, by Proposition \ref{p1.1'}.
   Denote by $\mu_0(dY_0)$ the Borel probability measure in ${\cal E}$
   giving the distribution of  $Y_0$.
  Without loss of generality, we may assume that
 $(\Omega,\Sigma,P)=({\cal E},{\cal B}({\cal E}),\mu_0)$
  and $Y_0(\omega,x)=\omega(x)$ for
  $\mu_0(d\omega)\times dx$-almost all
 $(\omega,x)\in{\cal E}\times\R^3$.
%%-----------------------------------------------
\medskip

Set ${\cal D}={\cal D}_0\oplus\R^3\oplus\R^3$,
${\cal D}_0:=[C_0^{\infty}(\R^3)]^2$, and
$$
\langle Y,Z\rangle:=\langle \phi,f\rangle+q\cdot u+p\cdot v \quad
\mbox{for }\,\,Y=(\phi,q,p)\in {\cal E}\quad\mbox{and }\,\,
Z=(f,u,v)\in {\cal D}.
$$
  For a probability  measure $\mu$ on ${\cal E}$,
 denote by $\hat\mu$ the characteristic functional (the Fourier transform)
$$
\hat\mu(Z)\equiv\int\exp(i\langle Y,Z\rangle )\,\mu(dY),\quad
 Z\in {\cal D}.
$$
A  measure $\mu$ is called {\em Gaussian} (with zero expectation)
if its characteristic functional is of the form
$\hat {\mu} (Z)=\exp\{-(1/2) {\cal Q}(Z,Z)\}$,
$Z\in {\cal D}$,
where ${\cal Q}$ is a  real nonnegative quadratic form
on ${\cal D}$.
A measure $\mu$ is called {\em translation-invariant} if
$\mu(T_h B)= \mu(B)$  for any $B\in{\cal B}({\cal E})$ and
$h\in\R^3$, where $T_h Y(x)= Y(x-h)$.

We assume that the initial measure $\mu_0$
has the following properties {\bf S0}--{\bf S3}.
\smallskip\\
{\bf S0} $\mu_0$ has zero expectation value, $\E
Y_0(x)\equiv\ds\int Y_0(x)\,\mu_0(dY_0)= 0$ for $x\in\R^3$.
\smallskip\\
{\bf S1} $\mu_0$ has finite mean energy density,
  i.e., $\E(|q_0|^2+|p_0|^2)<\infty$, and
\begin{equation}\label{med}
    \E\Big(|\varphi_0(x)|^2+|\nabla\varphi_0(x)|^2
+|\pi_0(x)|^2\Big) \le e_0<\infty.
  \end{equation}

Write $\mu^B_0:=P\mu_0$,  where
  $P:(\phi_0,q_0,p_0)\in {\cal E}\to\phi_0\in{\cal H}$.
 Now we impose conditions {\bf S2} and {\bf S3} on the
measure $\mu^B_0$. For simplicity of exposition, we assume that
$\mu^B_0$ has translation-invariant correlation matrices
(the case of non translation-invariant measures $\mu^B_0$ is considered in
Section \ref{sec8}). \smallskip \\
{\bf S2}  The correlation functions of the measure $\mu^B_0$,
$$
 Q^{ij}_0(x,y):=\int \phi^i_0(x)\phi^j_0(y)
 \,\mu_0^B(d\phi_0),
   \quad x,y\in\R^3,\quad \phi_0=(\phi^0_0,\phi^1_0)\equiv
   (\varphi_0,\pi_0),
$$
 are {\em translation-invariant}, i.e.,
$Q^{ij}_0(x,y)=q^{ij}_0(x-y)$, $i,j=0,1$.
%%%%%%%%%%%%%%%%%%%%%%%%%%%%%%%%%%%%%%%%%%%%%%%%%%%
\medskip

Now we formulate the {\em mixing condition} for the measure
$\mu^B_0$.

 Let ${\cal O}(r)$ be the set of all pairs of open convex
  subsets  ${\cal A}, {\cal B} \subset \R^3$ at distance
 $d({\cal A},\, {\cal B})\geq r$,
  and let $\sigma({\cal A})$ be the $\sigma $-algebra
     in ${\cal H}$ generated by
the linear functionals $\phi\mapsto\, \langle \phi,f\rangle$,
where  $f\in [C_0^\infty(\R^3)]^2$
 with $\supp f\subset {\cal A}$.
  Define the {\em Ibragimov mixing coefficient} of a probability
measure $\mu^B_0$ on ${\cal H}$ by the rule
(cf \cite[Def. 17.2.2]{IL})
  \beqn\label{ilc}
 \varphi(r)\equiv \sup_{({\cal A},{\cal B})\in {\cal O}(r)}
\sup_{\small\ba{c} A\in\sigma({\cal A}),B\in\sigma({\cal B})\\
\mu^B_0(B)>0\ea}
 \frac{|\mu^B_0(A\cap B)-\mu^B_0(A)\mu^B_0(B)|}{\mu^B_0(B)}.
    \eeqn
%----------------------------------------
\begin{definition}
 We say that the measure $\mu^B_0$ satisfies
the strong uniform Ibragimov mixing condition if
  $\varphi(r)\to 0$ as $r\to\infty$.
\end{definition}
%%%%---------------------------------------------------
{\bf S3} The measure $\mu^B _0$  satisfies the strong uniform
Ibragimov  mixing condition, and
\begin{equation}\label{1.12}
\int\limits_0^{+\infty} r^{d_F}\,\varphi^{1/2}(r)dr <\infty,
 \end{equation}
 where $d_F=d-1$ for the KGF, and $d_F=d-2$ for the WF, $d$ is dimension of the space.
%----------------------------------------
\begin{remark}
{\rm (i) The examples of the measures $\mu_0^B$ with zero mean
satisfying conditions (\ref{med}), {\bf S2} and {\bf S3} are given
in \cite[Section 2.6]{DKKS}.

 (ii) Instead of the {\it strong uniform} Ibragimov mixing
condition, it suffices to assume the {\it uniform} Rosenblatt
mixing condition \cite{Ros} together with a higher degree ($>2$)
in the bound (\ref{med}), i.e., to assume that there exists  a
$\delta$, $\delta >0$, such that
$$
\E\left(|\varphi_0(x)|^{2+\delta}+|\nabla\varphi_0(x)|^{2+\delta}
+|\pi_0(x)|^{2+\delta}\right) <\infty.
$$
In this case, the condition (\ref{1.12}) needs the following
modification: $\ds\int_0^{+\infty} \ds r^{d_F}\,\alpha^{p}(r)dr <\infty$,
where  $p=\min(\delta/(2+\delta), 1/2)$, $\alpha(r)$ is the Rosenblatt
mixing coefficient
 defined as in  (\ref{ilc}) but without $\mu^B_0(B)$ in the denominator.
}\end{remark}

%%%----------------------   2.4  ----------------
\subsection{Convergence to equilibrium for Klein-Gordon equations}
%%--------------------------------------------

We first consider the Cauchy problem for the wave (or
Klein--Gordon) equation,
\begin{equation}\label{KG-vc}
    \left\{\ba{l}
\ddot \varphi_t(x) =L_B\varphi_t(x),
 \quad t\in\R,\quad x\in\R^3,\\
\varphi_t(x)|_{t=0}=\varphi_0(x),\quad
\dot\varphi_t(x)|_{t=0}=\pi_0(x).
  \ea\right.
  \end{equation}
 %%---------------------------------------------------

Lemma \ref{lemma2.5} follows from \cite[Thms V.3.1, V.3.2]{Mikh}
as the speed of propagation for Eqn (\ref{KG-vc}) is finite.
\begin{lemma} \label{lemma2.5}
Let conditions {\bf A1}--{\bf A4} hold.
Then (i) for any $\phi_0=(\varphi_0,\pi_0)\in {\cal H}$,
 there exists a
unique solution $\phi_t=(\varphi_t(x),\dot\varphi_t(x))
\in C(\R,{\cal H})$ to the Cauchy problem (\ref{KG-vc}).\\
 (ii) For any $t\in \R$,
   the operator $W_t:\phi_0\mapsto\phi_t$
 is continuous on ${\cal H}$, and for any $T>0$, $R>R_a$,
 $$
 \sup_{|t|\le T}\Vert W_t\phi_0\Vert_R\le C(T)\Vert\phi_0\Vert_{R+T}.
 $$
\end{lemma}
%%-----------------------------------------------

 Let ${\cal E}_m(x)$ be the fundamental solution of the
operator  $-\Delta+m^2$, i.e.,
 $(-\Delta+m^2){\cal E}_m(x)=\delta(x)$.
  Since $d=3$,
 ${\cal E}_m(x)=e^{-m|x|}/(4\pi|x|)$.
 For almost all $x, y\in\R^3$, introduce the matrix-valued function
  $Q^B_\infty(x,y)=q^B_\infty(x-y)$,
   where
   \beqn\label{1.13}
q^B_{\infty}=
\frac{1}{2}\left(\ba{ll} q_0^{00}+{\cal E}_m * q_0^{11} &
q_0^{01}-q_0^{10} \\
 q_0^{10}-q_0^{01} & q_0^{11}+(-\Delta+m^2) q_0^{00}\ea\right).
   \eeqn
%%---------------------------------
Here $q^{ij}_{0}$, $i,j=0,1$, are correlation functions of
$\mu_0^B$ (see condition {\bf S2}),
$*$ stands  for the convolution.
%%--------------------------
We can rewrite $q^B_{\infty}$ in the Fourier transform as
\begin{equation}\label{2.12'}
 \hat q^B_{\infty}(k)
  = \frac{1}{2}\Big( \hat q_0(k)+\hat C(k)\hat q_0(k)
\hat C^T(k)\Big),
 \end{equation}
  where $(\,)^T$ denotes a matrix transposition, and
 \begin{equation}\label{C(k)}
  \hat C(k):=\left(
  \begin{array}{cc} 0 &\omega^{-1}(k) \\
-\omega(k)&0\end{array}\right),\quad \omega(k)=\sqrt{|k|^2+m^2}.
  \end{equation}
%%------------------------------------------
   \begin{remark}
{\rm
 Conditions {\bf S0}, (\ref{med}), {\bf S2} and {\bf S3} imply,
 by \cite[Lemma 17.2.3]{IL}, that   the derivatives
 $D^\alpha q_{0}^{ij}$ are bounded by the mixing coefficient:
$$
|D^\alpha q_{0}^{ij}(z)|\le Ce_0\,\varphi^{1/2}(|z|),
\quad \mbox{for any }\,\, z\in\R^3,\quad |\alpha|\le 2-i-j, \quad i,j=0,1.
$$
Therefore, $D^\alpha q_{0}^{ij}\in L^p(\R^3)$, $p\ge1$
(see \cite[p.16]{DKKS}). Hence, $(q^B_{\infty})^{ij}\in L^1(\R^3)$ if
$m\not=0$ by (\ref{1.13}).
If $m=0$, then the bound (\ref{1.12}) implies the
existence of the convolution
 ${\cal E}_m * q_{0}^{11}$ in (\ref{1.13}).
%%%%%%%%%%%%%%%%%%%%%%%------------------------------------------
%The matrix $\hat q_{\infty}(k)
% =\left(\hat q^{ij}_{\infty}(k)\right)_{i,j=0}^1$ satisfies
% the {\em equilibrium  condition}, i.e.,\\
% $\hat q^{11}_{\infty}(k)=\omega^2(k)\hat q^{00}_{\infty}(k)$,
%$\hat q^{01}_{\infty}(k)=-\hat q^{10}_{\infty}(k)$. Moreover,
%$\hat q^{00}_{\infty}(k)\ge0$, $\hat q^{01}_{\infty}(k)^*=-\hat
%q^{01}_{\infty}(k)$.
%%%%%%%%%%%%%%%%%%%%%%%%%%%%%%%%
 }  \end{remark}

 Denote by ${\cal Q}^{B,0}_{\infty}(f,f)$ the real quadratic form on
 ${\cal D}_0\equiv [C_0^\infty(\R^3)]^{2}$  defined by
\beqn\label{qpp}
  {\cal Q}^{B,0}_{\infty} (f,f)= \langle
Q^{B}_{\infty}(x,y),f(x)\otimes f(y)\rangle = \langle
q^B_{\infty}(x-y),f(x)\otimes f(y)\rangle.
  \eeqn
%-----------------------------------------------
\begin{definition}
 $\mu^B_t$ is a Borel probability measure in ${\cal H}$ which gives
the distribution of $\phi_t$:
$\mu^B_t(A)=\mu^B_0(W^{-1}_{t}A)$,
for any  $A\in {\cal B}({\cal H})$ and  $t\in \R$.
\end{definition}

For the measures $\mu^B_t$, the following result was proved in \cite{DK97}--\cite{DKRS}.

%%%%%%%%%%%%%%%%%%%%%%%%%%%%%%%%%%%%%%%%%%%%%%%%%%
\begin{theorem} \label{l6.2}
 Let conditions {\bf A1}--{\bf A4} hold and let the measure $\mu_0^B$ have zero mean
and satisfy conditions (\ref{med}), {\bf S2} and {\bf S3}. Then
(i) the measures $\mu^B_{t}$ %\equiv W_t^*\mu^B_0$
 weakly converge as
$t\to\infty$ on the space ${\cal H}^{-\ve}$ for each $\ve>0$.
 This means the convergence
 \begin{equation}\label{1.8}
 \int F(\phi)\mu^B_t(d\phi)\rightarrow
 \int F(\phi)\mu^B_\infty(d\phi)\quad as\quad t\to \infty
 \end{equation}
 for any bounded continuous functional $F(\phi)$
 on  ${\cal H}^{-\ve }$. \\
(ii) The limit measure  $\mu^B_{\infty}$
is a  Gaussian measure on ${\cal H}$.
The characteristic functional of $\mu^B_{\infty}$ is of the form
$\hat\mu^B_{\infty}(f)
=\exp\left\{-(1/2) {\cal Q}^{B}_{\infty}(f,f)\right\}$.
 Here
 \be\label{2.20}
{\cal Q}^{B}_{\infty}(f,f)={\cal Q}^{B,0}_\infty
(\Omega'f,\Omega' f), \quad f \in {\cal D}_0,
\ee
where $\Omega'$ is a linear continuous operator,
and $\Omega'=I$ in the case of the constant coefficients (see  Remark \ref{re2.9} below).\\
 (iii) The correlation matrices of $\mu_t^B$ converge to a limit, i.e.,
  for any $f_1,f_2\in {\cal D}_0$,
$$
  \int\langle \phi,f_1\rangle\langle \phi,f_2\rangle\,\mu_t^B(d\phi)
    \to{\cal Q}^B_{\infty}(f_1,f_2) \quad as\quad t\to\infty.
$$
  (iv) $\mu^B_{\infty}$ is invariant,
 i.e., $W_t^*\mu^B_{\infty}=\mu^B_{\infty}$, $t\in\R$.
 Moreover, the flow $W_t$ is mixing w.r.t. $\mu_\infty^B$, i.e.,
 the convergence (\ref{D.1}) holds.
  \end{theorem}
%%%%--------------------------------------------------
\begin{remark} \label{re2.9}
{\rm Now we explain the sense of the operator $\Omega'$ in (\ref{2.20}).
To prove (\ref{1.8}) in the case of variable coefficients,
we constructed in \cite{DKKS,DKRS} a version of the scattering theory
for solutions of infinite global energy.
%This allowed us to reduce the proof of (\ref{1.8}) to the case
%of constant coefficients.
Namely, in the case of the WF, we introduce appropriate  spaces
${\cal H}_{\gamma}$ of the initial data. By definition,
 ${\cal H}_{\gamma}$, $\gamma>0$, is the  Hilbert space
 of the functions $\phi=(\varphi,\pi)\in  {\cal H}$
with finite norm
$ \brr \phi\brr^2_{\gamma}=\int e^{-2\gamma|x|}
\Big(|\pi(x)|^2+|\nabla \varphi(x)|^2+|\varphi(x)|^2\Big)\,dx<\infty$.
It follows from (\ref{med})  that
  $\mu^B_0$ is concentrated in
${\cal H}_\gamma$ for all $\gamma>0$, since
\be\label{2.17}
\int\brr \phi_0\brr^2_{\gamma}\,\mu^B_0(d\phi_0)
\le e_0\int \exp({-2\gamma|x|})\,dx  <\infty.
\ee
Denote by $W_t$ the dynamical group of
Eqn (\ref{KG-vc}), while $W_t^0$ corresponds to
the 'free' equation, with $L_B=\Delta-m^2$.
In the WF case,
the following long-time asymptotics holds (see \cite{DKRS})
\begin{equation}\label{lt}
W_t\phi_0=\Omega W_t^0\phi_0+r_t\phi_0,\quad t>0,
  \end{equation}
where  $\Omega$ is a `scattering operator'.
$\Omega:\,{\cal H}_\gamma\rightarrow {\cal H}$
is a linear continuous operator for sufficiently small
$\gamma>0$.
The remainder $r_t$ is small in local energy seminorms
$\Vert\cdot\Vert_R$, $\forall R>0$:
$$
\Vert r_t\phi_0\Vert_R\to 0,\quad t\to\infty.
 $$
The representation (\ref{lt}) is based on our version
of the scattering theory for solutions of finite energy,
\begin{equation}\label{dsti}
(W_t)'f=(W_t^0)'\Omega' f+r'_t f,\quad t>0,
\end{equation}
where $(W_t)'$ and $(W^0_t)'$ are 'formal adjoint'
to the groups $W_t$ and $W^0_t$, respectively, see (\ref{2.25}).
$\Omega',r'_t:{\cal H}'\to{\cal H}'_\gamma$,
$\brr r'_tf\brr'_\gamma\to0$ as $t\to\infty$,
where $\brr\cdot\brr'_{\gamma}$ denotes the norm in the Hilbert space
${\cal H}'_{\gamma}$ dual to ${\cal H}_{\gamma}$. In particular,
for $f\in{\cal D}_0$, $\Omega'f\in{\cal H}'_\gamma$ and
the quadratic form ${\cal Q}^{B,0}_\infty$ from (\ref{2.20})
is continuous in ${\cal H}'_\gamma$
(for details, see Theorem 8.1 in \cite[p.1245]{DKRS}).

In the case of the KGF,
we derived in \cite{DKKS}
the dual representation (\ref{dsti}), where the remainder $r'_t$
is small in mean: $\E|\langle \phi_0,r'_t f\rangle|^2\to 0$, $t\to\infty$.
Moreover,
$\Omega'f\in H'_m\equiv L^2(\R^3)\oplus H^1(\R^3)$
for $f\in{\cal D}_0$,
and the quadratic form ${\cal Q}^{B,0}_\infty$
is continuous in $H'_m$.
%This version of scattering theory is essentially based on
%Vainberg's bounds \ci{V74,V89} for the local energy decay.
}
\end{remark}

%%%%%%%%%%%%%%------------------  2.5  ------------------------
\subsection{Convergence to equilibrium for the coupled system}
%%---------------------------------------------------------
To formulate the main result for the coupled system
  we introduce the following notations.
      Let $W'_t$ denote the  operator adjoint to $W_t$:
\be\label{2.25}
\langle \phi, W'_{t}f\rangle
  =\langle W_t\phi,f\rangle,\quad \mbox{for }\,
  f\in [S(\R^3)]^2, \quad  \phi\in{\cal H},\quad t\in\R.
\ee
%%-------------------

 Let $Z=(f,u,v)\in{\cal D}$, i.e.,
  $f\in [C_0^\infty(\R^3)]^2$, $(u,v)\in\R^3\times\R^3$. Write
\begin{equation}\label{hn}
\Pi(Z):=f_*(x) +\alpha(x)\cdot u+\beta(x)\cdot v.
  \end{equation}
  Here $\alpha=(\alpha_1,\alpha_2,\alpha_3)$,
 $\beta=(\beta_1,\beta_2,\beta_3)$,
 where
\beqn
\alpha_i(x) &:=& \ds \sum\limits_{r=1}^3
  \int\limits_0^{+\infty}
  {\cal N}_{ir}(s)W'_{-s}\nabla_r \rho_0\,ds,\quad
  \mbox{with }\,\rho_0:=(\rho,0),\label{al}\\
\beta_i(x)&:=&\ds \sum\limits_{r=1}^3\int\limits_0^{+\infty}
    {\cal N}_{ir}(s)W'_{-s}\nabla_r\rho^0\,ds,
  \quad \mbox{with }\,
 \rho^0:=(0,\rho),\quad i=1,2,3,  \label{beta}
 \eeqn
   the matrix-valued function
  ${\cal N}(s)=({\cal N}_{ir}(s))^3_{i,r=1}$, $s>0$,
  is  introduced in Corollary \ref{cor3.2}, and
\begin{equation}\label{teta}
f_*(x):=f(x)+\sum\limits_{i=1}^3\int\limits_0^{+\infty}
   \left(W'_{-s}\alpha_i\right)(x)
\left\langle W_s\nabla_i\rho^0,f\right\rangle\,ds.
 \end{equation}
%%%------------------------------

\begin{definition}
 $\mu_t$ is a Borel probability measure in ${\cal E}$ which gives
the distribution of $Y_t$: $\mu_t(B) = \mu_0(S^{-1}_{t}B)$,
$\forall B\in {\cal B}({\cal E})$,  $t\in \R$.
\end{definition}
%-----------------------------------------------

 Our main result is as follows.
%%%----------------------------
\begin{theorem}\label{tA}
Let conditions {\bf A1}--{\bf A5}, {\bf R1}--{\bf R3} and {\bf S0}--{\bf S3} hold. Then\\
(i) the measures $\mu_t$ weakly converge in the Fr\'echet spaces
  ${\cal E}^{-\ve}$ for each $\ve>0$,
\begin{equation}\label{convergence}
\mu_t\,\buildrel {\hspace{2mm}{\cal E}^{-\ve }}
 \over{-\hspace{-2mm} \rightharpoondown}\,
 \mu_\infty \quad as\quad t\to \infty,
  \end{equation}
  where $\mu_\infty$ is a limit measure on ${\cal E}$.
 This means the convergence
 $$
 \int F(Y)\mu_t(dY)\rightarrow
 \int F(Y)\mu_\infty(dY)\quad as\quad t\to \infty
 $$
 for any bounded continuous functional $F(Y)$
 on  ${\cal E}^{-\ve}$. \\
(ii) The limit measure $\mu_\infty$ is a Gaussian equilibrium
measure on ${\cal E}$.
 The limit characteristic functional is of the form
$\hat\mu_\infty(Z)=\exp\{-(1/2) {\cal Q}_\infty(Z,Z)\}$,
 $Z\in{\cal D}$. ${\cal Q}_{\infty} (Z,Z)$ denotes
  the real quadratic form on ${\cal D}$,
\begin{equation}\label{Qmu}
  {\cal Q}_{\infty} (Z,Z)={\cal Q}^B_{\infty} (\Pi(Z),\Pi(Z))=
{\cal Q}^{B,0}_{\infty} (\Omega'\Pi(Z),\Omega'\Pi(Z)),
 \end{equation}
  where ${\cal Q}^{B,0}_{\infty}$ is defined in (\ref{qpp}),
  and $\Pi(Z)$ is defined in (\ref{hn}).\\
(iii) The correlation functions of $\mu_t$ converge to a limit,
i.e., for any $Z_1,Z_2\in{\cal D}$,
\be\label{2.26}
\int \langle Y,Z_1\rangle\langle Y,Z_2\rangle\,\mu_t(dY)
 \to{\cal Q}_{\infty} (Z_1,Z_2)
 \quad as \,\,\,t\to\infty.
\ee
 (iv) The measure $\mu_\infty$ is
invariant, i.e., $S_t^* \mu_\infty=\mu_\infty$, $t\in\R$.\\
(v) The flow $S_t$ is mixing w.r.t.  $\mu_\infty$, i.e.,
$\forall F,G\in L^2({\cal E},\mu_\infty)$
 the following convergence holds,
$$
\lim_{t\to\infty} \int F(S_tY)G(Y)\mu_{\infty}(dY)
  =\int F(Y)\mu_{\infty}(dY)\int G(Y)\mu_{\infty}(dY).
$$
\end{theorem}
%%------------------------------------------------------

The assertions (i) and (ii) of Theorem \ref{tA} follow from
 Propositions \ref{l2.1} and \ref{l2.2} below.
%by using the same arguments as in \cite[Thm XII.5.2]{VF}.
%%%%%%%%%%%%%%---------------------------
  \begin{pro}\label{l2.1}
  The family of the measures $\{\mu_t,\, t\geq 0\}$
 is weakly compact in ${\cal E}^{-\ve}$ with any $\ve>0$.
 \end{pro}
%%%%%%%%%--------------------
  \begin{pro}\label{l2.2}
  For any $Z\in{\cal D}$,
$$
 \hat\mu_t(Z)\equiv
 \int\exp(i\langle Y,Z\rangle)\,\mu_t(dY)
\rightarrow \ds \exp\{-\frac{1}{2}{\cal Q}_\infty (Z,Z)\}, \quad
t\to\infty.
 $$
 \end{pro}

 Proposition \ref{l2.1} (Proposition \ref{l2.2})
 provides the existence (the uniqueness, resp.)
 of the limit measure $\mu_\infty$.
 Proposition \ref{l2.1} is proved in Section \ref{sec-com},
 Proposition \ref{l2.2} and the assertion~(iii)
 of Theorem~\ref{tA} are proved in Section \ref{sec-char}.
 Theorem \ref{tA} (iv) follows from (\ref{convergence})
 since the group $S_t$ is continuous in ${\cal E}$
  by Proposition \ref{p1.1'}~(ii).
  The assertion~(v) is proved in Section~7.
%%----------------------------------------------------

 %%---------------------------   3     ---------------------
\setcounter{equation}{0}
\section{Long-time behavior of the solutions}\label{sec3}
%%-------------------------------------------------
Using the operator $W_t$, we rewrite the system (\ref{1''}) in the
form
 \beqn
 \phi_t&=&W_t\phi_0+\int\limits_0^t q_s\cdot
W_{t-s}\nabla\rho^0\,ds,%\quad x\in\R^3,\quad t\in\R,
\label{cs1}\\
\ddot q_t&=&-\omega^2q_t+\langle\nabla\rho_0,\phi_t\rangle=
-\omega^2q_t+\int\limits_0^t D(t-s)q_s\,ds+F(t),\label{cs2}
 \eeqn
where $\phi_t=(\varphi_t(\cdot),\pi_t(\cdot))$,
 $\rho^0=(0,\rho)$, $\rho_0=(\rho,0)$,
  $F(t)$ denotes the vector-valued function,
  $ F(t)=\langle \nabla\rho_0,W_t\phi_0\rangle$,
  $D(t)$ stands for the matrix-valued function with entries
  \be\label{D}
  D_{ij}(t):=
   \langle\nabla_i\rho_0,W_t\nabla_j\rho^0\rangle,\quad i,j=1,2,3.
   \ee
   Note that in the case of the constant coefficients, i.e., $a_{ij}(x)\equiv \delta_{ij}$ and $a_0(x)\equiv0$ or $A_j(x)\equiv 0$,
 \be\label{3.4}
  D_{ij}(t)  %=\langle\nabla_i\rho_0,W_t^0\nabla_j\rho^0\rangle
    = (2\pi)^{-3}\int_{\R^3}
    k_ik_j\frac{\sin\omega(k)t}{\omega(k)}\,|\hat\rho(k)|^2\,dk,
    \quad \omega(k)=\sqrt{|k|^2+m^2},\quad m\ge0.
 \ee
%%%%------------------------

In sections~\ref{sec3} and \ref{sec5}, we study the long-time
behavior of the solutions $Y_t=(\phi_t,\xi_t)$ of problem
(\ref{1.1'}) by the following way. In Section~\ref{sec3.1}, we
prove the time decay for the solutions $q_t$ of (\ref{cs2}) with
$F(t)\equiv0$. Then
  we establish the time decay for the solutions $Y_t$
  of (\ref{1.1'}) in the case when the initial data of the field
  vanish for $|x|\ge R_0$ (Section~\ref{sec3.2}). Finally,
  for any initial data $Y_0\in{\cal E}$, we derive the long-time
  asymptotics of the solution $Y_t$ in the mean (Section~5).

  At first, consider the Cauchy problem
   for Eqn (\ref{cs2}) with $F(t)\equiv0$, i.e.,
   \beqn
  &&\ddot q_t=-\omega^2q_t+\int\limits_0^t D(t-s)q_s\,ds,
     \quad t>0,      \label{cs0}\\
&&q_t|_{t=0}=q_0,\quad \dot q_t|_{t=0}=p_0.\label{cs0-in}
  \eeqn
 %Eqn (\ref{cs0}) is so-called (vector)
 %{\it linear  Volterra integro-differential equation}.
 For the solutions of problem (\ref{cs0})--(\ref{cs0-in}),
 the following assertion holds.
%%-----------------------------
\begin{theorem}\label{h-eq}
  Let  conditions {\bf A1}--{\bf A5} and {\bf R1}--{\bf R3} be satisfied.
Then  %$q_t\in C^1(\R)$,
 $|q_t|+|\dot q_t|\le C\ve_F(t)(|q_0|+|p_0|)$   for any $t\ge0$.
  Here
\begin{equation}\label{ve(t)}
\ve_F(t)=\left\{\ba{ll} e^{-\delta t}\quad \mbox{with a }\,
\delta>0, & for\,\,the\,\, WF,\\
(1+t)^{-3/2}, & for\,\,the\,\,KGF.
  \ea\right.
  \end{equation}
\end{theorem}
%%----------------------------------------------------
\begin{cor}\label{cor3.2}
Denote by $V(t)$ a solving operator of the Cauchy problem
(\ref{cs0}), (\ref{cs0-in}). Then the variation constants formula
gives the following representation
for the solution of problem (\ref{cs2}), (\ref{cs0-in}):
$$
 \left(\ba{c}q_t\\ \dot q_t\ea\right)= V(t)
 \left(\ba{c}q_0\\ p_0\ea\right)+\int\limits_0^tV(s)
\left(\ba{c}0\\ F(t-s)\ea\right)\,ds,\quad t>0.
$$
Evidently, $V(0)=I$. The matrix
$V(t)$, $t>0$, is called the {\it resolvent} or
{\it principal matrix solution} for Eqn~(\ref{cs2}).
Theorem \ref{h-eq} implies that
$|V(t)|\le C\ve_F(t)$ with $\ve_F(t)$ from (\ref{ve(t)}), and for
 the solutions of  (\ref{cs2}) the following bound holds:
 \begin{equation}\label{est-cs}
 |q_t|+|\dot q_t|\le C_1\ve_F(t)(|q_0|+|p_0|)
   +C_2\int\limits_0^t\ve_F(s)|F(t-s)|\,ds, \quad \mbox{for }\,t\ge0.
 \end{equation}
Moreover, the matrix $V(t)$ has a form
$\left(\ba{cc}\dot{\cal N}(t)& {\cal N}(t)\\
\ddot {\cal N}(t)&\dot {\cal N}(t)\ea\right)$,
with matrix-valued entries satisfying the bound:
\be\label{NN}
|{\cal N}^{(j)}(t)|\le C\ve_F(t),\quad t>0,\quad j=0,1,2.\ee
\end{cor}
%%----------------------------------------------------

In next subsection, we prove Theorem~\ref{h-eq}
 for the WF case.
 %In this case, Theorem~\ref{h-eq} means that
 %the zero solution ($q_t\equiv0$) of Eqn (\ref{cs0})
 %is exponentially asymptotically stable.
  In the case of the KGF, Theorem~\ref{h-eq} can be proved
   combining the technique of \cite{IKV} and \cite[Appendix]{D10}, where Theorem~\ref{h-eq} was proved for the Klein-Gordon equation with constant coefficients,
   the methods of Section~\ref{sec3.1}, where
   the result is established in the case
   of the wave equations with variable coefficients, and Vainberg' results \cite{V74}
  for Klein-Gordon equations with variable coefficients.

%%%--------------------   3.1    ------------------
\subsection{Exponential stability of the zero solution
in the WF case} \label{sec3.1}
%%%%%%%%%%%%%%%%%

To prove Theorem \ref{h-eq}, we solve the Cauchy problem
(\ref{cs0}), (\ref{cs0-in}) by using the Laplace transform,
$$
\ti q(\lambda)=\ds\int\limits_0^{+\infty}
 e^{-\lambda t} q_t \, dt,\quad \Re \lambda>0.
$$
%Note that by (\ref{2.7}), $\ti Y(\lambda)$ is an analytic function in
%$\C_+=\{\lambda\in\C:\,\Re\lambda>0\}$ with values in $E$.
Then Eqn (\ref{cs0}) becomes
   \be\label{4.13'}
\lambda^2\ti q(\lambda)=-\omega^2\ti q(\lambda)+\tilde
D(\lambda)\ti q(\lambda)+p_0+\lambda q_0.
 \ee
 Let $H^s\equiv H^s(\R^3)$ denote the Sobolev space with norm $\Vert\cdot\Vert_s$.
Denote by $R_\lambda:H^0\to H^2$, $\Re\lambda>0$, an operator
such that $R_\lambda f=\varphi_\lambda(x)$ is a solution to the following equation
$$
(\lambda^2-L_B)\varphi_\lambda(x)=f(x).
$$
Then the entries of $\tilde D(\lambda)$ are
 \begin{equation}\label{4.9''}
\tilde D_{ij}(\lambda)
=\langle\nabla_i\rho,R_\lambda(\nabla_j\rho)\rangle,\quad i,j=1,2,3.
\end{equation}
Denote by $R^0_\lambda$, $\Re\lambda>0$,
 the operator $R_\lambda$ in the case when $L_B=\Delta$.
As shown in \cite[Lemma 3]{V69}, the operator
$R^0_\lambda(R_\lambda)$, $\Re\lambda>0$,
is analytic (finite-meromorphic, resp.) depends on $\lambda$.
By conditions~{\bf A1}--{\bf A3}, the operator $R_\lambda$, with $\Re\lambda>0$,
has not the poles and equals
$R_\lambda f= \int\limits_0^{+\infty}e^{-\lambda t}\varphi_t(x)\,dt$,
where $\varphi_t(x)$ is the solution to the Cauchy problem (\ref{KG-vc})
with initial data $\varphi_0\equiv0$,
$\pi_0=f\in H^0(\R^3)$.
Moreover, by energy estimates, the following bound holds
(see \cite[Theorem 2]{V69}),
\begin{equation}\label{3.9}
\Vert R_\lambda f\Vert_{1}+|\lambda|\Vert R_\lambda f\Vert_{0}
\le C\Vert f\Vert_0.
\end{equation}
We rewrite Eqn (\ref{4.13'}) as
$$
  \ti q(\lambda)=
  \left[(\lambda^2+\omega^2)I-\tilde D(\lambda)\right]^{-1}
 (p_0+\lambda q_0)
  \equiv \ti{\cal N}(\lambda)(p_0+\lambda q_0),
$$
  where $\ti {\cal N}(\lambda)$ stands for
  the $3\times 3$ matrix of the form
\begin{equation}\label{A}
 \ti{\cal N}(\lambda)=A^{-1}(\lambda),\,\,\mbox{with }\,\,
A(\lambda):=(\lambda^2+\omega^2)I-
\tilde D(\lambda) \quad\mbox{for }\,\, \Re\lambda>0.
  \end{equation}
%%---------------------------------------------------------

We first study properties of $A(\lambda)$. Write
$\C_\beta:=\{\lambda\in \C:\,\Re\lambda>\beta\}$
 for $\beta\in\R$.
%%%-------------------------------------------------
\begin{lemma}\label{detA}
Let  conditions {\bf A1}--{\bf A5} and {\bf R1}--{\bf R3} hold. Then\\
  (i)
$A(\lambda)$ admits an finite-meromorphic continuation to $\C$;
and there exists a $\delta>0$ such that
$A(\lambda)$ has not poles in $\C_{-\delta}$;\\
   (ii) for every $\beta\in(0,\delta)$, $\exists N_\beta>0$ such that
  $v\cdot A(\lambda)v\ge C|\lambda|^2|v|^2$
   for $\lambda\in \C_{-\beta}$
 with $|\lambda|\ge N_\beta$ and for every $v\in\R^3$.\\
   (iii)  There exists a $\delta_*>0$ such that
   $v\cdot A(\lambda)v\not=0$
     for $\lambda\in \overline\C_{-\delta_*}$ and for every
     $v\not=0$.
\end{lemma}
%----------------------------------------

In the case when $\rho(x)=\rho_r(|x|)$ and $L_B=\Delta$,
Lemma \ref{detA}  was proved in \cite[Lemma 7.2]{KSK}
(see also \cite[Lemma 4.3]{D10} in the case of the constant coefficients).
 \medskip\\
  %%------------------------------------------------------------
{\bf Proof}\,
Let $\psi$ be a smooth positive function which is like $e^{-|x|^2}$
as $|x|\to\infty$.
By $\hat R_\lambda$ ($\hat R^0_\lambda$) we denote the operator
$R_\lambda$ ($R^0_\lambda$, resp.)
which is considered as an operator from $H^0_b$ to $H^1_\psi$,
where $H_b^s=\{f\in H^s:f(x)=0 \,\,\mbox{for } |x|\ge b\}$ with a norm
$\Vert\cdot\Vert_{s,b}$,
$H^s_\psi$ is the space with a norm $\Vert\varphi\Vert_{s,\psi}=
\Vert\psi\varphi\Vert_s$.
We choose a $b$ such that $b\ge\max\{R_\rho,R_a\}$
(see conditions {\bf A2} and {\bf R2}).
\medskip

Now we state properties ({\bf V1})--({\bf V4})
of the operator $\hat R_\lambda$ which follow from
 Vainberg's results \cite{V69,V89}.\\
({\bf V1}) (see \cite[Theorem 3]{V69})
The operator $\hat R^0_\lambda$
admits an analytic continuation on  $\C$,
and for any $\gamma>0$,
$$
\Vert \hat R_\lambda^0 f\Vert_{1,\psi}+
|\lambda|\Vert \hat R_\lambda^0 f\Vert_{0,\psi}\le
C(\gamma)\Vert f\Vert_{0,b},\quad |\Re \lambda|<\gamma.
$$
The operator $\hat R_\lambda$
admits a finite-meromorphic continuation on  $\C$,
and for any $\gamma>0$ there exists $N=N(\gamma)$ such that
in the region $M_{\gamma,N}:=\{\lambda\in\C:\,
|\Re\lambda|\le\gamma,\,|\Im\lambda|\ge N\}$
the following estimate holds:
$\Vert \hat R_\lambda f\Vert_{j,\psi}\le
2\Vert \hat R_\lambda^0 f\Vert_{j,\psi}$, $j=0,1$,
for $f\in H^0_b$ (see \cite[Theorem 4]{V69}).\\
({\bf V2})  For any $\gamma>0$, $\hat R_\lambda$
has at most a finite number of poles in the domain $\C_{-\gamma}$.\\
({\bf V3})
$\hat R_\lambda$ has not poles for $\Re\lambda\ge0$,
by conditions {\bf A1}--{\bf A3}.\\
({\bf V4}) There exist constants $C,T,\alpha,\beta>0$ such that
for any $f\in H^0_b$,
$$
\Vert\hat R_\lambda f\Vert_{0,\psi}\le C|\lambda|^{-1}
e^{T|\Re\lambda|}\Vert f\Vert_{0,b},\quad
\mbox{for }\,\lambda\in U_{\alpha,\beta}=\{\lambda\in\C:\,
|\Re\lambda|<\alpha\ln|\Im\lambda|-\beta\}.
$$

We return to the proof of Lemma \ref{detA}.

(i) In the case of the constant coefficients, i.e., when $L_B=\Delta$,
  $\tilde D_{ij}(\lambda)
  =\langle\nabla_i\rho,R^0_\lambda(\nabla_j\rho)\rangle$
  admits an analytic continuation to $\C$.
  Therefore, in this case,
  $A(\lambda)$ admits an analytic continuation to $\C$.
  In the general case, item~(i) of Lemma~\ref{detA}
  follows from  ({\bf V1})--({\bf V3}).

(ii) By (\ref{4.9''}) and (\ref{3.9}), $\tilde D_{ij}(\lambda)\to0$
as $|\lambda|\to\infty$ with $\Re\lambda>0$. On the other hand,
property~({\bf V1}) implies that for any $\gamma>0$
there exists $N=N(\gamma)>0$ such that
\begin{equation}\label{3.12}
   |\tilde D_{ij}(\lambda)|\le C(\gamma)|\lambda|^{-1}
   \quad\mbox{for } \,\, \lambda\in M_{\gamma,N}.
\end{equation}
   Hence,  there exists a $\beta>0$ such that
 $|\tilde D_{ij}(\lambda)|\le C|\lambda|^{-1}\to0$
 as $|\lambda|\to\infty$ with $\lambda\in\C_{-\beta}$.
This implies the assertion (ii) of Lemma~\ref{detA}.
\smallskip

  (iii)  Note first that  $\det A(\lambda)\not=0$ for $\Re\lambda>0$,
  by (\ref{2.7}).
    Further, the matrix $A(\lambda)$ is positive definite
for $\Im\lambda=0$.
  Indeed, let $\lambda=\mu\in\R\setminus 0$, and
  put $f=\nabla\rho\cdot v$ with $v\in\R^3$.
     Then  $f\in H^0_b$ and $\hat f|k|^{-1}\in H^0$.
     Denoting $\varphi_\mu= R_\mu f\in H^2$, we obtain
$$
   \langle f,R_\mu f\rangle
   =\langle \varphi_\mu,(\mu^2-L_B)\varphi_\mu\rangle
   \ge \alpha
   \Vert\nabla \varphi_\mu \Vert^2_0
   =\alpha\Vert\nabla(R_\mu f)\Vert^2_0,
  $$
  by condition {\bf A3}. On the other hand,
  $\langle f,R_\mu f\rangle\le \Vert\nabla(R_\mu f)\Vert_0\cdot
  \Vert F^{-1}(|k|^{-1}\hat f)\Vert_0$.
  Hence,
 $$
  \Vert\nabla(R_\mu f)\Vert_0\le \frac{1}{\alpha}
  \Vert F^{-1}(|k|^{-1}\hat f)\Vert_0.
  $$
   Therefore, for any $\mu>0$ and $v\in\R^3\setminus\{0\}$, we obtain
  \beqn\label{3.13}
   v\cdot \tilde D(\mu)v&=& \langle f,R_\mu f\rangle \le
   \frac{1}{\alpha} \left\Vert F^{-1}\left(|k|^{-1}\hat f\right)\right\Vert^2_0
    =\frac{1}{\alpha(2\pi)^{3}}\left\Vert |k|^{-1}\hat f\right\Vert^2_0\nonumber\\
    &=&
 \frac{1}{\alpha(2\pi)^{3}}\int\frac{(k\cdot v)^2}{|k|^2}|\hat\rho(k)|^2\,dk
    %=\frac{1}{\alpha}v\cdot K_0v
    <\omega^2|v|^2,
  \eeqn
   by condition~{\bf R1}.
%%-------------------------------------------------
  In the case $\mu=0$, put $\hat R_0f:=\lim_{\ve\to+0} \hat R_{\ve}f$,
  where the limit is understood in the space $H^1_\psi$.
  Then
% $|\langle f, \hat R_\ve f\rangle|<\omega^2|v|^2$ for any $\ve>0$, then
$|\langle f, \hat R_0 f\rangle|<\omega^2|v|^2$ by (\ref{3.13}).
    Hence, for any $v\in\R^3\setminus0$ and $\mu\in\R$,
$$
 v\cdot A(\mu)v=(\mu^2+\omega^2)|v|^2
  -v\cdot \tilde D(\mu)v
  %\langle \nabla\rho\cdot v,\hat R_\mu(\nabla\rho\cdot v)\rangle
  %\ge\omega^2|v|^2-v\cdot K_0 v
  >0.
    $$
%%------------------------------------------------
Moreover, there exists a $\delta_0$, $\delta_0>0$, such that
$$
v\cdot A(\lambda)v\not=0\quad
\mbox{for }\,\,|\lambda|<\delta_0\quad
\mbox{and for any }\,\,v\in\R^3\setminus\{0\}.
$$
%%-----------------------------------------------------
Now let $\lambda=iy+0$ with $y\in\R$,
 and put again $f=\nabla\rho\cdot v\in H^0_b$.
 By property ({\bf V1}), there exists $N_0>0$ such that
$v\cdot A(iy)v\sim(\omega^2-y^2)|v|^2+C|v|^2/|y|\not=0$
for   $|y|\ge N_0$ and $v\not=0$.
Hence, to prove the assertion (iii) of Lemma \ref{detA},
it suffices to show that
$$
 \det A(iy+0)=\det \left[(\omega^2-y^2)I-\tilde D(iy+0)\right]
 \not=0\quad \mbox{for }\,\,\delta_0\le|y|\le N_0.
$$
%%---------------------
 In \cite{D10}, we have proved that in the case when $L_B=\Delta$,
condition {\bf R3} and the Plemelj formula~\cite{GSh} yield
\begin{equation}\label{3.15}
 \Im\langle f, \hat R^0_{iy+0}f\rangle
 =-\frac\pi2 y^3(2\pi)^{-3}\int\limits_{|\theta|=1}
 (v\cdot \theta)^2|\hat\rho(|y|\theta)|^2\,dS_\theta\not=0\quad
 \mbox{for any }\,v,y\in\R^3\setminus\{0\},
 \end{equation}
where $\hat R^0_{iy+0}f:=\lim_{\ve\to+0} \hat R^0_{iy+\ve}f$.
In the case when $L_B=L_W$, we can choose $M_a$ so small that
\begin{equation}\label{3.22}
v\cdot \Im\tilde D(iy+0)v\equiv\Im\langle f, \hat R_{iy+0}f\rangle\not=0
\quad\mbox{for all }\,\,v\in\R^3\setminus\{0\}
\quad\mbox{and }\,\,|y|\in(\delta_0, N_0)
\end{equation}
(see condition {\bf A5}).
In fact, we split $\langle f, \hat R_{iy+0}f\rangle$
 into two terms
\be\label{split}
\langle f, \hat R_{iy+0}f\rangle=
\langle f, \hat R^0_{iy+0}f\rangle+
\langle f, (\hat R_{iy+0}-\hat R^0_{iy+0})f\rangle.
\ee
Since $f=\nabla\rho\cdot v$,
then there exists a constant $C_0>0$ such that
for $|y|\in(\delta_0, N_0)$ we have
\be\label{addterm}
|\langle f, (\hat R_{iy+0}-\hat R^0_{iy+0})f\rangle|=
|\langle f, \hat R_{iy+0}(L_B-\Delta)\hat R^0_{iy+0}f\rangle|
\le C_0\Vert \rho\Vert_1^2\,|v|^2M_a,
\ee
%%%Moreover,
%%%$\sup_y\Vert\hat R_{iy+0}f\Vert_{0,b}\le C_1\Vert f\Vert_{0,b}$,
%%%$\sup_y\Vert\hat R^0_{iy+0}f\Vert_{0,b}\le C_0\Vert f\Vert_{0,b}$.
%%--------------------------------------------------------------
where $M_a=\max_{x\in\R^3}\{|a_{ij}(x)-\delta_{ij}|,|a_0(x)|\}$.
Hence, (\ref{3.15}) and (\ref{addterm}) imply that
if $M_a$ is enough small, then
(\ref{3.22}) holds. For example, assume that
$$
M_a\le\frac{M}{2C_0\Vert\rho\Vert^2_1}, \quad\mbox{with }\,\,
M=\min_{\delta_0\le|y|\le N_0}
\min_{v\not=0}\frac{|v\cdot S(y)v|}{|v|^2}>0,
$$
where $S(y)$, $y\in\R^3$, stands for the $3\times3$ matrix with the entries
$S_{ij}(y)$,
$$
S_{ij}(y)=
  \frac\pi2 y^3(2\pi)^{-3}\int\limits_{|\theta|=1}
  \theta_i\theta_j|\hat\rho(|y|\theta)|^2\,dS_\theta,\quad
  i,j=1,2,3.
$$
Hence, for $|y|\in(\delta_0,N_0)$,
$|\Im\langle f, \hat R^0_{iy+0}f\rangle|
=|v\cdot S(y)v|\ge 2C_0\Vert\rho\Vert^2_1\, |v|^2 M_a$.
Therefore, (\ref{split}) and (\ref{addterm}) imply  bound (\ref{3.22}).
Finally,
   $v\cdot\Im A(iy+0)v=-v\cdot\Im \tilde D(iy+0)v\not=0$.
   Therefore, there exists   $\delta_*>0$ such that
   $v\cdot \Im A(iy+x)v\not=0$ for $|x|\le\delta_*$.
Lemma \ref{detA} is proved. \bo
%%-------------------------------------------------
\medskip

For any $\delta<\delta_*$,
denote by ${\cal N}(t)$ the inverse Laplace transformation of
$\ti{\cal N}(\lambda)$,
$$
{\cal N}(t)=\frac{1}{2\pi i}
\int\limits_{-i\infty-\delta}^{i\infty-\delta}
e^{\lambda t}\ti{\cal N}(\lambda)\,d\lambda,\quad \,t>0.
$$
%%----------------------------------------------------
\begin{lemma}\label{c1}
Let $L_B=L_W$ and conditions {\bf A1}--{\bf A5}, {\bf R1}--{\bf R3} hold. Then,
  for $j=0,1,\dots$ and any $\delta<\delta_*$,
 \begin{equation}\label{decayA}
    |{\cal N}^{(j)}(t)|\le C e^{-\delta t},\quad t>1.
   \end{equation}
\end{lemma}
%%--------------------------
{\bf Proof}\, By Lemma \ref{detA},
 the bound on ${\cal N}(t)$ follows.
 To prove the bound for $\dot{\cal N}(t)$, we
consider $\lambda\ti{\cal N}(\lambda)$ and prove the bound
\begin{equation}\label{4.14}
  \left|v\cdot(\lambda\ti{\cal N}(\lambda))'v\right|
\le\frac{C|v|^2}{1+|\lambda|^{2}} \quad \mbox{for }\,
\lambda\in\overline\C_{-\delta}.
  \end{equation}
   Therefore,
$$
\left|t\dot{\cal N}(t)\right|=C
  \left|\int\limits_{\Re\lambda=-\delta}
 e^{\lambda t}(\lambda\ti{\cal N}(\lambda))'\,d\lambda
\right|\le C_1e^{-\delta t},
$$
  and bound (\ref{decayA}) for $\dot{\cal N}(t)$
follows. By Lemma \ref{detA} (ii), to prove bound (\ref{4.14}),
  it suffices to show that
 $|\ti{\cal N}'_{ij}(\lambda)|\le C(1+|\lambda|)^{-3}$.
    Since $R'_\lambda f=-2\lambda R^2_\lambda f$, then
    by formulas (\ref{4.9''}), (\ref{3.9}) and property ({\bf V1}), we have
$$
|\tilde D'_{ij}(\lambda)|\le
  2|\lambda||\langle \nabla_i\rho,R_\lambda^2 \nabla_j\rho\rangle|
  \le C<\infty \quad   \mbox{as }\,\,|\lambda|\to\infty
\quad \mbox{with }\quad \lambda\in\C_{-\delta}.
$$
Therefore, (\ref{A}) and Lemma \ref{detA} imply that, for $i,j=1,2,3$,
$$
|\ti{\cal N}'_{ij}(\lambda)|%\le \frac{C}{1+|\lambda|^4}
%\left(\sum\limits_{i,j=1}^d|A'_{ij}(\lambda)|\right)
 \le
\frac{C_1}{1+|\lambda|^{3}}\quad \mbox{as }\,|\lambda|\to\infty.
$$
This yields (\ref{4.14}). Bound (\ref{decayA})
  with $j\ge2$  can be proved in a similar way.
  \bo
%%--------------------------------------------
\begin{cor}
The solution of the Cauchy problem (\ref{cs0})--(\ref{cs0-in})
 is $q_t= \dot{\cal N}(t)q_0+{\cal N}(t)p_0$. Therefore,
in the case of the WF, Lemma \ref{c1} implies Theorem~\ref{h-eq} with any $\delta<\delta_*$.
\end{cor}

%%-------------------------  3.2   -------------------
\subsection{ Time decay for $Y_t$ when
$\phi_0(x)=0$ for $|x|\ge R_0$}\label{sec3.2}
%%----------------------------

For the solution $Y_t$ of (\ref{1.1'}), the following bound holds.
%%-------------------------------------------------
\begin{lemma}\label{l5.1}
Let conditions {\bf A1}--{\bf A5} and {\bf R1}--{\bf R3}
hold and let $Y_0\in {\cal E}$
be such that
\begin{equation}\label{**}
  \varphi_0(x)=\pi_0(x)=0\quad \mbox{for }\,|x|>R_0,
\end{equation}
    with some $R_0>0$.
 Then for every $R>0$ there exists a constant $C=C(R,R_0)>0$ such that
\begin{equation}\label{delocen}
  \Vert Y_t\Vert_{{\cal E},R} \le C\ve_F(t)\Vert
Y_0\Vert_{{\cal E},R_0},\quad t\ge0.
   \end{equation}
Here $\ve_F(t)=(1+t)^{-3/2}$ for the KGF. In the case of the WF,
 $\ve_F(t)=e^{-\delta t}$
  with a $\delta\in(0,\min(\delta_*,\gamma))$, where
   constants $\delta_*$  and $\gamma$
   are introduced in Lemma \ref{detA} (iii)
 and in bound (\ref{freewave}), respectively.
\end{lemma}
%%---------------------------
 %In the case when $m=0$, Lemma \ref{l5.1}
%is an extension of Proposition 7.1 in \cite{KSK}, where a
%  similar result was established for $a(x)\equiv0$ and
%$\rho(x)=\rho_r(|x|)$.
%%-------------------------------------------
{\bf Proof}\, Step (i): At first, we prove bound
  (\ref{delocen}) for $\xi_t=(q_t,p_t)$. In the case of the WF, condition (\ref{**}) and the Vainberg bounds
 (see \cite{V89} or \cite[Proposition 10.1]{DKRS}) imply that,
   for any $R>0$,
   there exist constants $\gamma=\gamma(R,R_0)>0$
    and $C=C(R,R_0)>0$ such that
   \be\label{freewave}
 \Vert W_t\phi_0\Vert_R\le
 Ce^{-\gamma t}\Vert\phi_0\Vert_{R_0},\quad t\ge0.
   \ee
 Therefore, bound (\ref{est-cs}) with
 $F(t)\equiv\langle \nabla\rho_0,W_t\phi_0\rangle =
    -\langle \rho_0,\nabla W_t\phi_0\rangle$
    and condition~{\bf R2} yield
\be\label{3.21}
|\xi_t|\le C_1e^{-\delta t}|\xi_0|
   + C(\rho)\int\limits_0^t e^{-\delta s}
 \Vert \nabla(W_{t-s}\phi_0)^0\Vert_{L^2(B_{R_\rho})}\,ds
  \le C e^{-\delta t} \Vert Y_0\Vert_{{\cal E},R_0},
 \ee
with any $\delta<\min(\delta_*,\gamma)$.
If $L_B=L_{KG}$, then we apply the Vainberg bound \cite{V74}:
\be \label{KG}
   \Vert W_t\phi_0\Vert_R \le C
(1+t)^{-3/2}\Vert\phi_0\Vert_{R_0},\quad t\ge0.
 \ee
Hence, $|F(t)|\le C(1+t)^{-3/2}\Vert\phi_0\Vert_{R_0}$, and
 bound (\ref{delocen}) for $\xi_t$ follows from (\ref{est-cs}).
\medskip

Step (ii): Now we prove bound (\ref{delocen}) for $\phi_t$.
  In the case of the WF, Eqn (\ref{cs1}),  condition
(\ref{**}), bounds (\ref{freewave}) and (\ref{3.21}) yield
$$
 \Vert\phi_t\Vert_R
  \le C_1e^{-\gamma t}\Vert\phi_0\Vert_{R_0}
   +C_2\int\limits_{0}^t e^{-\delta s} \Vert
Y_0\Vert_{{\cal E},R_0}e^{-\gamma(t-s)}\,ds
\le Ce^{-\delta t} \Vert Y_0\Vert_{{\cal E},R_0},\quad t\ge0,
  $$
 with any $\delta<\min(\delta_*,\gamma)$.
 For the KGF,  the bound
  $ \Vert\phi_t\Vert_R
%\le C(1+t)^{-3/2}\Vert \phi^0\Vert_{R_1}+C\int\limits_{0}^t
%(1+t-s)^{-3/2}(1+s)^{-3/2}\,ds\Vert Y_0\Vert_{{\cal E},R_1}\nonumber\\
\le  C(1+t)^{-3/2}\Vert Y_0\Vert_{{\cal E},R_0}$
follows from Eqn~(\ref{cs1}), bound (\ref{KG}),
  and estimate (\ref{delocen}) for $q_t$.
 This proves Lemma \ref{l5.1}.
 \bo

%%%%%%%%%%%%%%%%%%%%%%%%%%%%% 4   %%%%%%%%%%
\setcounter{equation}{0}
\section{Compactness of the measures $\mu_t$}\label{sec-com}
%%%%%%%%%%%%%%%%%%%%%%%%%%%%%%%%%%%%%%%%%
Proposition \ref{l2.1}  can be deduced from bound (\ref{7.1.1})
below by the Prokhorov Theorem \cite[Lemma II.3.1]{VF} using the
method of \cite[Theorem XII.5.2]{VF}, since the embedding
${\cal E}\equiv {\cal E}^0\subset {\cal E}^{-\ve}$
is compact for every $\ve>0$.

\begin{lemma}
Let conditions {\bf A1}--{\bf A5}, {\bf R1}--{\bf R3}
and {\bf S0}--{\bf S2} hold. Then
  \be\label{7.1.1}
   \sup\limits_{t\ge 0} \E\Vert S_t
Y_0\Vert^2_{{\cal E},R} \le C(R)<\infty,\quad \forall R>0.
\ee
\end{lemma}
{\bf Proof}\, Let $\rho\equiv0$. In this case,
we denote by $S^0_t$ the solving operator $S_t$.
Note first that
   \be\label{7.1.10}
  \sup\limits_{t\ge 0}
  \E\Vert S_t^0 Y_0\Vert^2_{{\cal E},R} \le C(R),
  \quad\forall R>0.
  \ee
Indeed, by the notation (\ref{2.1'}),
$
\Vert S_t^0 Y_0\Vert^2_{{\cal E},R}= \Vert W_t
\phi_0\Vert^2_{R}+|q_t^0|^2+|\dot q_t^0|^2,
$
where $q_t^0$ is a solution to the Cauchy problem
$$
\ddot q_t^0+\omega^2 q_t^0=0,\quad t\in\R,\quad
 (q_t^0,\dot q_t^0)|_{t=0}=(q_0,p_0).
$$
Hence, $|q_t^0|+|\dot q_t^0|\le C(|q_0|+|p_0|)$.
 By \cite[bound (11.2)]{DKKS}
 and \cite[bound (9.2)]{DKRS},
  we have
   \be\label{6.3}
 \sup\limits_{t\in\R}\E\Vert W_t \phi_0\Vert^2_{R}\le C(R),
 \quad \forall R>0.
  \ee
  This implies (\ref{7.1.10}).
  Further, we represent the solution to problem (\ref{1.1'}) as
 $$
  S_tY_0=S_t^0Y_0+\int\limits_0^t S_{t-\tau} B S^0_\tau Y_0\,d\tau,
 $$
  where, by definition,
$BY=
\left(0,0,q\cdot\nabla\rho,\langle\varphi,\nabla\rho\rangle\right)$
for $Y=(\varphi,q,\pi,p)$.
 Hence, condition {\bf A2}, (\ref{delocen}), and
(\ref{7.1.10}) yield
$$
   \ba{ccl} \E\Vert S_t Y_0\Vert^2_{{\cal E},R}
&\le& \E\Vert S_t^0
  Y_0\Vert^2_{{\cal E},R} +\E\int\limits_0^t
\Vert S_{t-\tau} B S^0_\tau Y_0\Vert^2_{{\cal E},R}\,d\tau\\
 &\le&
C(R)+\int\limits_0^t \ve^2_F(t-\tau)\,
\E\Vert S^0_\tau Y_0\Vert^2_{{\cal E},R_\rho}\,d\tau
\le C_1(R)<\infty.\,\,\,\bo \ea
 $$

%%%%%%%%%%%%%%%%%%%%%%%%%%%%% 5  %%%%%%%%%%%%%%%%%%%%%%%%%%%%%%%
\setcounter{equation}{0}
\section{Asymptotic behavior for $Y_t=(\phi_t,q_t,p_t)$ in mean}
\label{sec5}
%%%%%%%%%%%%%%%%%%%%%%%%%%%%%%%%%%%%%%%%%%%%%%%%%%%%%%%%%%%%%%%

%%---------------------------------------
\begin{pro}\label{l7.1}
Let  conditions {\bf A1}--{\bf A5}, {\bf R1}--{\bf R3}
 and {\bf S0}--{\bf S2} be satisfied.

(i) The following bounds hold,
  \beqn\label{8.1}
\E|q_t-\langle W_t\phi_0,\alpha\rangle|^2
&\le&C\ti\ve_F(t),\\
 \E|p_t-\langle W_t\phi_0,\beta \rangle|^2&\le &
C\tilde\ve_F(t),\quad t>0,     \label{8.2}
  \eeqn
   where  the functions $\alpha$ and $\beta$
 are defined in (\ref{al}) and (\ref{beta}),
 $\ti\ve_F(t)= (1+t)^{-1}$ for the KGF, and
  $\ti\ve_F(t)= \ve^2_F(t)=e^{-2\delta t}$
  with a $\delta>0$ for the WF.
\medskip

(ii) Let $f\in [C_0^\infty(\R^3)]^2$ with $\supp f\subset B_R$.
    Then, for $t\ge1$,
   \be\label{8.3}
\E\Big|\langle\phi_t,f\rangle - \left\langle
W_t\phi_0,f_*\right\rangle\Big|^2 \le C\ti\ve_F(t),
  \ee
  where the function $f_*$ is defined in (\ref{teta}).
\end{pro}
%%%%%%%%%%%%%%%%%%%%%%%%%%%%%%%%%%%%%%%%%%%
{\bf Proof}\, (i) At first, Theorem~\ref{h-eq} and
Corollary \ref{cor3.2} yield
  \be\label{8.4}
  \E \Big|q_t-\int\limits_0^t {\cal N}(s)
  \left\langle W_{t-s}\phi_0,
\nabla\rho_0 \right\rangle\,ds\Big|^2
 \le C\ve^2_F(t)
  \ee
 with $\ve_F(t)$ from (\ref{ve(t)}).
  Further,
  \beqn
 &&\E\left|\int\limits_t^{+\infty}{\cal N}_{ir}(s)
\langle W_{t-s}\phi_0,\nabla_r\rho_0 \rangle\,ds\right|^2
 \nonumber\\
  &=& \int\limits_t^{+\infty}\! {\cal N}_{ir}(s_1)\,ds_1
  \int\limits_t^{+\infty}\!{\cal N}_{ir}(s_2) \E\Big(
\langle W_{t-s_1}\phi_0,\nabla_r\rho_0\rangle
  \langle W_{t-s_2}\phi_0,\nabla_r\rho_0\rangle\Big)\,ds_2.
\nonumber
\eeqn
 For any $t,s_1, s_2\in\R$,
   \beqn
  \Big|\E\Big(\langle W_{t-s_1}\phi_0,\nabla_r\rho_0\rangle
\langle W_{t-s_2}\phi_0,\nabla_r\rho_0\rangle\Big)\Big|
   &\le& C \sup_{\tau\in\R}
  \E|\langle W_{\tau}\phi_0,\nabla_r\rho_0\rangle|^2
\nonumber\\
&\le& C_1\sup_{\tau\in\R}
 \E\Vert W_{\tau}\phi_0\Vert^2_{R_\rho}
  \le C_2<\infty\nonumber
  \eeqn
 by  bound (\ref{6.3}).
   Hence, using (\ref{NN}), we obtain
   \be\label{8.7}
\E\left|\int\limits_t^{+\infty} {\cal N}(s)
   \langle W_{t-s}\phi_0,
   \nabla\rho_0\rangle \,ds\right|^2
   \le \left(\int\limits_t^{+\infty}
   \ve_F(s)\,ds\right)^2=C_1\,\ti\ve_F(t).
 \ee
Therefore, (\ref{8.1}) follows from (\ref{8.4}), (\ref{8.7}) and
(\ref{al}) because
 $$
   \left\langle W_{t-s}\phi_0,\nabla\rho_0\right\rangle
 = \left\langle W_t\phi_0,W'_{-s}\nabla\rho_0\right\rangle.
 $$
  The bound (\ref{8.2}) can be proved in a similar way.
\medskip
%%-----------------------

(ii) Let $f\in [C_0^\infty(\R^3)]^2$ with $\supp f\subset B_R$.
  By  Eqn (\ref{cs1}), we have
 \beqn\label{6.7}
   \langle \phi_t,f\rangle
  = \langle W_t\phi_0,f\rangle+
\int\limits_0^{t} q_{t-s}\cdot
  \left\langle W_s \nabla\rho^0,f\right\rangle\,ds.
 \eeqn
   Using Vainberg's bounds \cite{V74,V89}, we obtain
\beqn\label{6.8}
   \langle W_s \nabla\rho^0,f\rangle=\left\{
\ba{ll} {\cal O}(e^{-\gamma |s|})\quad \mbox{with a }\,\gamma>0
&\mbox{if }\, L_B=L_W,\\
{\cal O}((1+|s|)^{-3/2})&\mbox{if }\, L_B=L_{KG}. \ea\right.
  \eeqn
 If $L_B=L_W$ we put
  $\tilde \ve_F(t)=\ve^2_F(t)=e^{-2\delta t}$ with any $\delta<\min(\delta_*,\gamma)$, see Lemma \ref{l5.1}.
   Applying the Parseval inequality and
   bounds (\ref{8.1}) and (\ref{6.8}), we get
\beqn\label{7.8}
\E\Big|\int\limits_0^{t}
  \Big(q_{t-s}- \langle W_{t-s}\phi_0,\alpha\rangle\Big)\cdot
 \left\langle W_s\nabla\rho^0, f\right\rangle \,ds\Big|^2
%%%-----------------------------------------------------------
%\nonumber\\
%&\le& \int\limits_0^{t}
%  (\E\left|q_{t-s}-\langle W_{t-s}\phi_0,
%  \alpha\rangle\right|^2)^{1/2}
%  |\left\langle W_s\nabla\rho^0,\psi\right\rangle|\,ds
% \int\limits_0^{t}
%  (\E\left|q_{t-s}-\langle W_{t-s}\phi_0,
%  \alpha\rangle\right|^2)^{1/2}
%  |\left\langle W_s\nabla\rho^0, f\right\rangle|\,ds\nonumber\\
%%%-------------------------------------------------------
 &\le&\left(\int_0^t(\tilde\ve_F(t-s))^{1/2}
  |\left\langle W_s\nabla\rho^0,f\right\rangle|\,ds\right)^2
    \nonumber\\
      &\le& C\ti\ve_F(t).
  \eeqn
 %   Indeed, for $m\not=0$, by direct calculation,
%$$\int_0^t(\tilde\ve_m(t-s))^{1/2}
%  |\left\langle W_s\nabla\rho^0,f\right\rangle|\,ds
%  \le C \int_0^t(1+t-s)^{-1/2}
%  (1+s)^{-3/2}\,ds\le C_1(1+t)^{-1/2}.
%  $$

   Write
 $  I(t):= \E\Big|\int\limits_t^\infty
 \langle W_{t-s}\phi_0,\alpha\rangle\cdot
\langle W_s\nabla\rho^0,f\rangle\,ds\Big|^2$.
  Then
  \be\label{6.9}
  |I(t)|\le C\ti\ve_F(t).
   \ee
 This follows from (\ref{6.8}) and from
  the following estimate:
$$
  \E|\langle W_\tau\phi_0,\alpha\rangle|^2
   =\sum\limits_{i=1}^3\E\Big| \sum\limits_{r=1}^3\int\limits_0^{+\infty}
{\cal N}_{ir}(s)\left\langle W_{\tau-s}\phi_0, \nabla_r\rho_0
\right\rangle\,ds\Big|^2 \le C<\infty,
  \quad \mbox{for }\,\,\tau\in\R,
   $$
    by (\ref{6.3}) and (\ref{NN}).
Relation (\ref{6.7}) and bounds (\ref{7.8}) and (\ref{6.9}) imply
 (\ref{8.3}). \bo
%%----------------------------------------------------------
\begin{cor}\label{c7.2}
Let $Z=(f,u,v)\in {\cal D}=[C_0^\infty(\R^3)]^2
\times\R^3\times\R^3$.
   Then
$$ \langle Y_t,Z\rangle=
 \langle W_t \phi_0, \Pi(Z)\rangle + r(t),
$$
   where $\Pi(Z)$ is defined in (\ref{hn}),
  $\langle Y_t,Z\rangle=\langle\phi_t, f\rangle +
q_t \cdot u+p_t\cdot v$,
   $Y_t=(\phi_t,q_t,p_t)$ is a solution
to the Cauchy problem (\ref{1.1'}),
 and $\E\,\left(|r(t)|^2\right)\le C \ti\ve_F(t)$.
\end{cor}
%%--------------------------------------------

%%------------------------------  6 ------------------------
\setcounter{equation}{0}
\section{Convergence of characteristic
functionals and correlation functions}\label{sec-char}
%%-------------------------------------------------------

{\bf Proof of Proposition \ref{l2.2}}\, By the triangle
inequality,
  \beqn\label{8.16}
   \left|\E e^{i\langle Y_t,Z\rangle}-
e^{-\frac{1}{2} {\cal Q}_\infty(Z,Z)}\right| \le\Big|\E
\Big(e^{i\langle Y_t,Z\rangle}-
 e^{i\langle W_t\phi_0,\Pi(Z)\rangle}\Big)\Big|
+ \Big|\E e^{i\langle W_t\phi_0,\Pi(Z)\rangle}
-e^{-\frac{1}{2}{\cal Q}_\infty (Z,Z)}\Big|.
 \eeqn
 Applying Corollary \ref{c7.2}, we estimate
 the first term in the r.h.s. of (\ref{8.16}) by
$$
 %  \Big|\E \Big(e^{i\langle Y_t,Z\rangle}- e^{i\langle
%W_t\phi_0,\Pi(Z)\rangle}\Big)\Big| \le
  \E\Big|\langle Y_t,Z\rangle
-\langle W_t\phi_0,\Pi(Z)\rangle\Big|
 \le \E|r(t)| \le
\Big(\E|r(t)|^2\Big)^{1/2} \le C \ti\ve^{1/2}_m(t)\to0\quad
\mbox{as }\,t\to\infty,
  $$
   It remains to prove the convergence
$\E\left(\exp\{i\langle W_t\phi_0,\Pi(Z)\rangle\}\right)
  \equiv\hat\mu^B_t(\Pi(Z))$ to a limit
as $t\to\infty$.

  In \cite{DKKS, DKRS}, we
have proved the convergence of $\hat\mu^B_t(f)$ to a limit for
$f\in{\cal D}_0\equiv[C_0^\infty(\R^3)]^2$. However,
$\Pi(Z)\not\in{\cal D}_0$ in general.
Consider the cases of the WF and KGF separately.

In the WF case,
$\Pi(Z)\in{\cal H}'_\gamma$ if $Z\in{\cal D}$,
for sufficiently small $\gamma>0$,
where ${\cal H}'_\gamma$ is introduced in Remark~\ref{re2.9}.
This follows from formulas
(\ref{hn})--(\ref{teta}), from the bound (\ref{NN}), and from the estimate
$$
\brr W'_tf\brr'_\gamma\le Ce^{\gamma|t|}\brr f\brr'_\gamma,\quad t\in\R,
\quad \mbox{for any }\,\,f\in{\cal H}'_\gamma.
$$
The last estimate can be proved in a similar way as the same estimate for $(W_t^0)'$ in \cite[lemma 8.2]{DKRS} using the energy estimates.
%%------------------------------------------------------
\begin{lemma}\label{l-m=0}
 Let $L_B=L_W$. Then the quadratic forms ${\cal Q}^{B}_t(f,f):=
\ds\int|\langle\phi_0,f\rangle|^2\mu^B_t(d\phi_0)$,
    $t\in\R$, and the characteristic functionals $\hat\mu^B_t(f)$,
$t\in\R$, are equicontinuous on ${\cal H}'_\gamma$.
\end{lemma}
%%--------------------------------------------------
{\bf Proof}\, In the case when  $L_B=\Delta$,
Lemma \ref{l-m=0} was proved in \cite[Corollary 4.3]{DKRS}.
In the general case, i.e., when $L_B=L_W$,
this lemma can be proved by a similar way
and the proof is based on the bound
$\E\brr W_t\phi_0\brr^2_\gamma\le C<\infty$
for any $\gamma>0$. Now we prove this bound.
By (\ref{6.3}), we have
\be\label{6.10}
e_t:=\E(|\varphi_t(x)|^2+|\nabla\varphi_t(x)|^2+|\pi_t(x)|^2)
\le C<\infty,
\ee
since $\E\Vert W_t\phi_0\Vert_R^2=e_t|B_R|$,
where $|B_R|$ denotes the volume of the ball $B_R=\{x\in\R^3:|x|\le R\}$.
Hence, the bound (\ref{6.10}) implies, similarly to (\ref{2.17}), that
for any $\gamma>0$ there is a constant $C=C(\gamma)>0$ such that
$$
\E\brr W_t\phi_0\brr^2_{\gamma}
=e_t\int \exp({-2\gamma|x|})\,dx \le C <\infty. \,\,\,\,\bo
$$

In the case of KGF, we write $H'_m=L^2(\R^3)\oplus H^1(\R^3)$.
Then $\Pi(Z)\in H'_m$ if $Z\in {\cal D}$.
 This follows from formulas
(\ref{hn})--(\ref{teta}) and the bound (\ref{NN}).
%%-----------------------------------------------------------
\begin{lemma}\label{l}
Let $L_B=L_{KG}$. Then
(i) the quadratic forms ${\cal Q}^B_t(f,f)$,
    $t\in\R$, are equicontinuous on $H'_m$,
(ii) the characteristic functionals $\hat\mu^B_t(f)$,
$t\in\R$, are equicontinuous on $H'_m$.
\end{lemma}
{\bf Proof}\, (i) It suffices to prove the uniform bound
\be\label{6.12}
  \sup\limits_{t\in\R} |{\cal Q}^B_t(f,f)|\le C
\Vert f\Vert_{H'_m}^2\quad \mbox{for any }\,\,f\in H'_m.
  \ee
At first, note that
  $ {\cal Q}^B_t(f,f)
 =\langle Q_{0}(x,y), W'_t f(x)\otimes W'_t f(y)\rangle$.
%%-----------------
On the other hand, by conditions {\bf S0}, {\bf S2} and {\bf S3}, the
correlation functions $Q_0^{ij}(x,y)$ of the measure $\mu_0^B$
satisfy the following bound:
 for $\alpha,\beta\in\Z^3$, $|\alpha|\le 1-i$,
$|\beta|\le 1-j$, $i,j=0,1,$
 \be\label{6.3'}
 |D_{x,y}^{\alpha,\beta}Q^{ij}_0(x,y)| \le
Ce_0\varphi^{1/2}(|x-y|),\quad x,y\in\R^3,
 \ee
 according to \cite[Lemma 17.2.3]{IL}. Therefore, by (\ref{1.12}),
$$
 \int\limits_{\R^3}
|D_{x,y}^{\alpha,\beta}Q^{ij}_0(x,y)|^p\,dy\le Ce_0^p
\int\limits_{\R^3} \varphi^{p/2}(|x-y|)\,dy\le
 C_1e_0^p
 \int\limits_0^\infty r^{2}\varphi^{1/2}(r)dr
   <\infty,\quad p\ge1.
$$
  Hence, by the Shur lemma,  the quadratic form
   $\langle Q_0(x,y),f(x)\otimes f(y)\rangle$ is
continuous in $[L^2(\R^3)]^2$.
% Moreover,  the quadratic form ${\cal Q}^B_\infty(f,f)$ is also
%  continuous in $[L^2(\R^3)]^2$.
%%----------------------------
Therefore,
$$
\sup\limits_{t\in\R} |{\cal Q}^B_t(f,f)|
=\sup\limits_{t\in\R}|\langle Q_{0}(x,y), W'_t f(x)\otimes W'_t f(y)\rangle|\le C
\sup\limits_{t\in\R} \Vert W'_t f\Vert_{L^2}^2\le C\Vert
f\Vert_{H'_m}^2.
$$
The last inequality follows from the energy conservation
for the Klein--Gordon equation.\\
%%%---------------------------------
 (ii) By the Cauchy-Schwartz inequality and (\ref{6.12}),
   we obtain
$$
\ba{rcl}
  \left|\hat\mu^B_t(f_1)-\hat\mu^B_t(f_2)\right|
  &=&\left|\ds\int\Big(e^{i\langle \phi_0,f_1\rangle}
  -e^{i\langle \phi_0,f_2\rangle}\Big)\mu^B_t(d\phi_0)\right|
 \le\ds\int
\left|e^{i\langle\phi_0,f_1-f_2\rangle}-1\right|
\mu^B_t(d\phi_0)\\
&\le&
  \ds\int|\langle \phi_0,f_1-f_2\rangle|\mu^B_t(d\phi_0)
\le\sqrt{\ds\int|\langle \phi_0,f_1-f_2 \rangle|^2
\mu^B_t(d\phi_0)}\\
&=&\sqrt{{\cal Q}^B_t(f_1-f_2,f_1-f_2)}
    \le C\Vert f_1-f_2 \Vert_{H'_m}.~~~~~\bo \ea
$$
%%-----------------------------------

We return to the proof of Proposition \ref{l2.2}.
By \cite[Proposition 2.3]{DKS} (or \cite[Proposition~3.3]{DKM2}),
and by Lemmas \ref{l-m=0} and \ref{l}, the characteristic functionals
$\hat\mu^B_t(\Pi(Z))$ converge to a limit
as $t\to\infty$.
  This completes the proof of Proposition \ref{l2.2}
  and Theorem \ref{tA},  (i)--(ii). \bo
%%--------------------------------------------------

\begin{lemma}
Let all assumptions of Theorem \ref{tA} be satisfied. Then
convergence (\ref{2.26}) holds.
%% for $Z_1,Z_2\in {\cal D}$,
%%  \be\label{concorf}
%%   \E\left(\langle Y_t,Z_1\rangle\langle Y_t,Z_2\rangle\right)
%%   \to {\cal Q}_\infty(Z_1,Z_2),\quad t\to\infty,
%%   \ee
%% where ${\cal Q}_{\infty}(Z,Z)$ is defined in (\ref{Qmu}).
\end{lemma}
%%--------------------------------------------
 {\bf Proof}\, It suffices to prove
the convergence of
  $\int|\langle Y,Z\rangle|^2\,\mu_t(dY)=\E|\langle Y_t,Z\rangle|^2$
  to a limit as $t\to\infty$.
  It follows  from Corollary \ref{c7.2} that
   for $Z\in {\cal D}$,
$$
  \E|\langle Y_t,Z\rangle|^2=
  \E|\langle W_t\phi_0,\Pi(Z)\rangle|^2+o(1)
   ={\cal Q}^B_t(\Pi(Z),\Pi(Z))+o(1),
   \quad t\to\infty,
$$
where $\Pi(Z)$ is defined in (\ref{hn}).
 Therefore, by the results from \cite{DKS,DKM2}
  and by Lemmas \ref{l-m=0} and \ref{l}, the quadratic forms
${\cal Q}^B_t(\Pi(Z),\Pi(Z))$ converge to a limit as $t\to\infty$.
 Formula (\ref{Qmu}) implies (\ref{2.26}).
 \bo

\newpage
%%%%%%%%%%%%%%%%%%%%%%%%%%%  7  %%%%%%%%%%%%%%%%%%555
\setcounter{equation}{0}
\section{Ergodicity and mixing for the limit measures}\label{sec7}
%%%%%%%%%%%%%%%%%%%%%%%%%%%%%%%%%%%%%%%%%%%%%%%%%%%
Denote by $\E_\infty$ ($\E^B_\infty$) the integral w.r.t.
$\mu_\infty$ ($\mu^B_\infty$, respectively). In \cite{DK97}, we have
proved that $W_t$ is mixing w.r.t. $\mu^B_\infty$, i.e., for any
$f,g\in L_2({\cal H},\mu^B_\infty)$, the following convergence
holds,
 \be\label{D.1}
 \E_\infty^B\left(f(W_t\phi)g(\phi)\right)\to
  \E_\infty^B (f(\phi)) \,\E_\infty^B  (g(\phi)) \quad \mbox{as }\,\,t\to\infty.
\ee

  Recall that the limit measure $\mu_\infty$ is invariant by
Theorem \ref{tA} (iv). Now we prove that the
flow $S_t$ is mixing w.r.t. $\mu_\infty$.
 This mixing property %for $S_t$
means that
  the convergence (\ref{convergence}) holds for the initial measures
  $\mu_0$ that are absolutely continuous w.r.t. $\mu_\infty$,
  and the limit measure coincides with $\mu_\infty$.
 %%------------------------------------------------
\begin{theorem}\label{tD}
The phase flow $S_t$ is mixing w.r.t. $\mu_\infty$, i.e., for any
 $F,G\in L_2({\cal E},\mu_\infty)$ we have
$$
\E_\infty\left(F(S_tY)G(Y)\right)\to
  \E_\infty (F(Y))\, \E_\infty (G(Y))\quad \mbox{as }\,\,t\to\infty.
$$
In particular, the flow $S_t$ is ergodic w.r.t. $\mu_\infty$,
i.e., for any $F\in L_2({\cal E},\mu_\infty)$,
$$
\lim_{T\to\infty}\frac1T\int\limits_0^T F(S_tY)\,dt
 =\E_\infty(F(Y))\quad(\mod\mu_\infty).
$$
\end{theorem}

 To prove Theorem \ref{tD}, we introduce new notations.
Represent $Y\in{\cal E}$ as $Y=(Y^0,Y^1)$ with
$Y^0=(\varphi(\cdot),q)\in H^1_{\rm loc}(\R^d)\times\R^d$,
$Y^1=(\pi(\cdot),p)\in L^2_{\rm loc}(\R^d)\times\R^d$,
 and  $Z\in{\cal D}$ as $Z=(Z^0,Z^1)$ with $Z^0=(f^0(\cdot),u^0)$,
$Z^1=(f^1(\cdot),u^1)\in C_0^\infty(\R^d)\times\R^d$. For
$t\in\R$, introduce a "formal adjoint" operator $S'_t$ on the
space ${\cal D}$ by the rule
  \be\label{D.2}
 \langle S_tY, Z\rangle=\langle Y,S'_t Z\rangle,
 \quad Y\in{\cal E},\quad Z\in{\cal D}.
\ee
%%------------------------------------------------
\begin{lemma}
For $Z\in{\cal D}$,
 \be\label{D.3}
  S'_tZ    =(\dot f_t(\cdot),\dot u_t,f_t(\cdot),u_t),
   \ee
where $(f_t(x),u_t)$ is the solution of system (\ref{1''}) with
the initial data (see (\ref{5''})) \\
 $(\varphi_0,q_0,\pi_0,p_0)=(f^1,u^1, f^0,u^0)$.
\end{lemma}
{\bf Proof}\, Differentiating (\ref{D.2}) in $t$ with
 $Y,Z\in{\cal D}$, we obtain
$\langle \dot S_tY, Z\rangle=\langle Y,\dot S'_t Z\rangle$.
 The group $S_t$ has the generator
 \be\label{7.4}
  {\cal L}=\left(\ba{cc}0&1\\{\cal A}&0\ea\right),
   \quad \mbox{with }\,\,
 {\cal A}\left(\ba{c}\varphi\\q\ea\right)
  =\left(\ba{c} L_B\varphi+q\cdot \nabla\rho\\
 -\omega^2 q+\langle\nabla\rho,\varphi\rangle\ea\right).
 \ee
 The generator of $S'_t$ is the conjugate operator
${\cal L}'=\left(\ba{cc}0&{\cal A}\\1&0\ea\right)$. Hence,
(\ref{D.3}) holds with
  $\left(\ba{c}\ddot f_t(x)\\\ddot u_t\ea\right)
   ={\cal A}\left(\ba{c} f_t(x)\\u_t\ea\right)$.\bo\\
   \medskip
%%-----------------------------------------------
 % \begin{cor}
 %   Let $Z\in{\cal D}$, with  $f^0(x)=f^1(x)=0$ for $|x|\ge R_0$.
 % Then, by Lemma \ref{l5.1}, the following bound holds,
 % $ \Vert S'_t Z\Vert_{{\cal E},R}
 %  \le C\ve_m(t)\Vert Z\Vert_{{\cal E},R_0}. $
 %\end{cor}
 %%%%-----------------------------------------------------

 Since the limit measure $\mu_\infty$ is Gaussian with zero mean,
 the proof of Theorem \ref{tD} reduces to that of the following
 convergence.
 \begin{lemma}\label{lD}
 For any $Z_1,Z_2\in{\cal D}$,
\be\label{D.4}
  \E_\infty\Big(\langle S_tY, Z_1\rangle
  \langle Y, Z_2\rangle\Big)\to0,\quad t\to\infty.
\ee
\end{lemma}
{\bf Proof }\, First we note that, by relation (\ref{Qmu}),
$$
\E_\infty\Big(\langle Y, Z_1\rangle
  \langle Y, Z_2\rangle\Big)=
\E^B_\infty\Big(\langle \phi, \Pi(Z_1)\rangle
  \langle \phi, \Pi(Z_2)\rangle\Big),
$$
where $\Pi(Z)$ is defined in (\ref{hn}). Secondly, for fixed
$t$, we have $S'_tZ\in{\cal D}$. Further,
  \beqn\label{D.5}
 \E_\infty\Big(\langle S_tY, Z_1\rangle
  \langle Y, Z_2\rangle\Big)\!\!\!&=&\!\!\!
  \E_\infty\Big(\langle Y, S'_tZ_1\rangle
  \langle Y, Z_2\rangle\Big)=
  \E^B_\infty\Big(\langle \phi, \Pi(S'_tZ_1)\rangle
  \langle \phi, \Pi(Z_2)\rangle\Big)\nonumber\\
  \!\!\!&=&\!\!\!\E^B_\infty\Big(\langle \phi, (\Pi S'_t-W'_t\Pi)Z_1\rangle
  \langle \phi, \Pi(Z_2)\rangle\Big)+
\E^B_\infty\Big(\langle \phi, W'_t\Pi(Z_1)\rangle
  \langle \phi, \Pi(Z_2)\rangle\Big)\nonumber\\
  \!\!\!&=&\!\!\!I_1+I_2.
    \eeqn
Note that
$\langle \phi, \Pi(Z)\rangle\in L_2({\cal H},\mu^B_\infty)$
 for all $Z\in{\cal D}$. Indeed, by (\ref{Qmu}),
$$
\E_\infty^B|\langle \phi, \Pi(Z)\rangle|^2
 ={\cal Q}^B_\infty(\Pi(Z),\Pi(Z))
={\cal Q}_\infty(Z, Z)<\infty.
$$
Therefore, the convergence (\ref{D.1}) implies
 \beqn
 I_2&\equiv&\E^B_\infty\Big(\langle \phi, W'_t\Pi(Z_1)\rangle
  \langle \phi, \Pi(Z_2)\rangle\Big)\nonumber\\
  &=&\E^B_\infty\Big(\langle W_t\phi, \Pi(Z_1)\rangle
  \langle \phi, \Pi(Z_2)\rangle\Big)\to
  \E_\infty^B \Big(\langle \phi, \Pi(Z_1)\rangle\Big)
    \E_\infty^B \Big(\langle \phi, \Pi(Z_2)\rangle\Big),
    \quad t\to\infty.\nonumber
\eeqn
 On the other hand,
$\E_\infty^B \langle \phi, \Pi(Z_i)\rangle
 = \E_\infty \langle Y, Z_i\rangle=0$,
 for $Z_i\in{\cal D}$,
because $\mu_\infty$ has zero mean. Therefore,
 \be\label{D.6}
I_2\to0,\quad t\to\infty.
  \ee
Now we prove that
$\E_\infty^B\left|\langle\phi,
\Pi(S'_tZ)-W'_t\Pi (Z)\rangle\right|^2=0$ for all $t>0$.
 This yields
   \be\label{D.7}
 I_1\equiv \E^B_\infty\Big(\langle \phi,
\Pi (S'_t Z_1)-W'_t\Pi(Z_1)\rangle
  \langle \phi, \Pi(Z_2)\rangle\Big)=0.
\ee
 Indeed, by Corollary \ref{c7.2},
$$
\E|\langle S_{\tau+t}Y,Z\rangle-
 \langle W_{\tau+t}\phi,\Pi(Z)\rangle|^2\to0,\quad \tau\to\infty.
$$
On the other hand, since
  $\langle S_{\tau+t}Y,Z\rangle=\langle
S_{\tau}Y,S'_{t}Z\rangle$, we have, for all $t>0$,
$$
\E|\langle S_{\tau}Y,S'_{t}Z\rangle-\langle
W_{\tau}\phi,\Pi(S'_tZ)\rangle|^2\to0,\quad \tau\to\infty.
$$
Therefore, by the triangle inequality,
$$
A:=\E|\langle W_{\tau}\phi,\Pi(S'_tZ)\rangle - \langle
W_{\tau+t}\,\phi,\Pi(Z)\rangle|^2\to0,\quad \tau\to\infty.
$$
Since
  $\langle W_{\tau+t}\,\phi,\Pi(Z)\rangle
  =\langle W_{\tau}\,\phi,W'_{t}\Pi(Z)\rangle$,
  we obtain
$$
A=\E|\langle W_{\tau}\phi,
 \Pi(S'_tZ)-W'_{t}\Pi(Z)\rangle|^2\to0,
  \quad \tau\to\infty.
 $$
Hence, by Theorem \ref{l6.2} (iii) and Lemmas \ref{l-m=0} and \ref{l},
$$
\E_\infty^B|\langle\phi,\Pi(S'_tZ)-W'_{t}\Pi(Z)\rangle|^2
=\lim_{\tau\to\infty}
 \E|\langle W_{\tau}\phi,\Pi(S'_tZ)-W'_{t}\Pi(Z)\rangle|^2
  =0 \quad \mbox{for all }\,\,t>0.
 $$
  Finally, (\ref{D.5})--(\ref{D.7}) imply the convergence
  (\ref{D.4}). Theorem \ref{tD} is  proved. \bo

%%%%%%%%%%%%%%%%%%%%%%%%%%%%%%%  8  %%%%%%%%%%%%%%%%%%%%%%%%%%%
\setcounter{equation}{0}
\section{Non translation invariant initial measures}\label{sec8}
%----------------------------------

In this section we extend the results of Theorem \ref{tA} to the case of
 non translation-invariant initial measures. Note that the
proof of Theorem \ref{tA} is based on two assertions.
 We first derive the asymptotic behavior of solutions  $Y_t$ in mean:
 $\langle Y_t,Z\rangle\sim\langle W_t\phi_0,\Pi(Z)\rangle$
  as $t\to\infty$ (see Corollary \ref{c7.2}).
This asymptotics allows us to reduce the convergence analysis for
the coupled system to the same problem for the  wave (or
Klein-Gordon) equation.
 The second assertion is the weak convergence of the measures $\mu_t^B=W_t^*\mu_0^B$ to a limit
 as $t\to\infty$ (see Theorem~\ref{l6.2}).
However, the weak convergence of $\mu_t^B$ holds under weaker
conditions on $\mu_0^B$ than {\bf S2} and {\bf S3}.
  Now we formulate these conditions
  (see \cite{DKS} for $L_B=L_W$ and \cite{DKM2} for $L_B=L_{KG}$).

%%----------------------------- 8.1  --------------
   \subsection{ Conditions on $\mu_0^B$}\label{sec8.1}
%%----------------------------------

 In the case of the KGF,  we assume  that $\mu_0^B$ has zero mean,
 satisfies a mixing condition {\bf S3} and has
 a finite mean energy density
 (see (\ref{med})), i.e.,
 \beqn\label{med'}
 \int\Big(|\varphi_0(x)|^2+|\nabla\varphi_0(x)|^2+|\pi_0(x)|^2\Big)
 \mu_0^B(d\phi_0)\!\!&=&\!\!Q_0^{00}(x,x)
 +[\nabla_x\nabla_y Q_0^{00}(x,y)]\Big|_{x=y}+
 Q_0^{11}(x,x)\nonumber\\&  \!\!\le&\!\! e_0<\infty.
 \eeqn
 However, condition {\bf S2} of translation invariance for $\mu_0^B$
  can be weakened as  follows.\\
 {\bf S2'}\, The correlation functions of the measure $\mu^B_0$
     have   the form
   \be\label{q1}
Q^{ij}_0(x,y)= q^{ij}_-(x-y)\zeta_-(x_1)\zeta_-(y_1)+
q^{ij}_+(x-y) \zeta_+(x_1)\zeta_+(y_1),\quad i,j=0,1.
  \ee
 Here  $q^{ij}_{\pm}(x-y)$ are the correlation functions of some
translation-invariant measures $\mu^B_{\pm}$
  with zero mean value in ${\cal H}$,
   $x=(x_1,\dots,x_d)$, $ y=(y_1,\dots,y_d)\in\R^d$,
 the functions $\zeta_\pm\in C^{\infty}(\R)$ such that
  \be\label{zeta}
  \zeta_{\pm}(s)= \left\{
\ba{ll}
1,~~\mbox{for }~ \pm s>\,a,\\
0,~~\mbox{for }~ \pm s<-a,\ea\right.
 \ee
and $a>0$. The measure  $\mu^B_0$ is not translation-invariant if
$q^{ij}_-\not=q^{ij}_+$.
%%%%%%%%%%%%%%%%%%%%%%%%%%%%%%%%%%%
\medskip

 In the case of WF,  instead of {\bf S2} and {\bf S3}
we impose the following conditions {\bf S2'} and {\bf S3'}.\\
 {\bf S2'}\, The correlation functions of  $\mu^B_0$ have
the form
  \beqn\label{3}
    Q^{ij}_0(x,y) = \left\{\ba{l}
q^{ij}_-(x-y),\quad x_1, y_1<\!\!-a,\\
 q^{ij}_+(x-y),\quad x_1, y_1>\,a,\ea \right.
  \eeqn
  with some $a>0$ and $q^{ij}_\pm$ as in (\ref{q1}).
  However, in the WF case, instead
of (\ref{med'}) we impose a stronger condition (\ref{H2}).
   Namely, the following derivatives are
continuous and the bounds hold,
   \beqn\label{H2}
|D_{x,y}^{\alpha,\beta} Q^{ij}_{0}(x,y)|\le \left\{ \ba{l}
C\nu_\kappa(|x-y|)\quad\mbox{if }\,\,\kappa=0,1,\dots, d-2\\
C\nu_{d-1}(|x-y|)\quad \mbox{if }\,\,\kappa=d-1,d,d+1
 \ea\right|\,\,\,
\kappa=i+j+|\alpha|+|\beta|,
  \eeqn
  with $|\alpha|\le (d-3)/2+i$, $|\beta|\le (d-3)/2+j$, $i,j=0,1$.
Here  $\nu_\kappa\in C[0,\infty)$ ($\kappa=0,\dots,d-1$) denote
some continuous nonnegative nonincreasing functions in
$[0,\infty)$
 with the finite integrals
 $\ds\int\limits_0^{\infty}(1+r)^{\kappa-1}\nu_\kappa(r) dr<\infty$.
 Moreover, for $d\ge5$,
$\ds\int\limits_0^{\infty}(1+r)^{d-4+\kappa}\nu_\kappa(r)dr<\infty$
with $\kappa=0,2$.\\
  %%%%%%%%%%%%%%%%%%%%%%%%%%%%%%%%%%%%%%%%%%%%%%%%%%%
%Now we formulate a new mixing condition {\bf S3'} for the measure
%$\mu^B_0$ (if $m=0$).\\
%%------------------------------------------------------------
{\bf S3'}\, Let ${\cal O}(r)$ be the set of all pairs of open convex
  subsets  ${\cal A}, {\cal B} \subset \R^d$ at distance
 $d({\cal A},\, {\cal B})\geq r$,  and let
 $\alpha=(\alpha_1,\dots,\alpha_d)$ with integers $\alpha_i\ge 0$.
Denote by $\sigma_{i\alpha} ( {\cal A})$
 the $\sigma$-algebra of the subsets in ${\cal H}$ generated by all
  linear functionals
$$\phi_0=(\phi_0^0,\phi_0^1)\mapsto
 \langle D^{\alpha}\phi_0^i,f\rangle,\quad \mbox{with }\,\,
|\alpha|\le 1-i,\quad i=0,1, $$
  where $f\in C_0^\infty(\R^d)$
 with $ \supp f\subset {\cal A}$.
For $\kappa=0,1$, let $\sigma_\kappa({\cal A})$ be the
$\sigma$-algebra generated by $\sigma_{i\alpha}({\cal A})$ with
$i+|\alpha|\ge \kappa$, i.e., $\sigma_\kappa({\cal A})\equiv
\bigvee\limits_{i+|\alpha|\ge \kappa}\sigma_{i\alpha}({\cal A})$.
 We define the (Ibragimov)
mixing coefficient of  $\mu^B_0$ on
 ${\cal H}$ as (cf. (\ref{ilc}))
$$
   \varphi_{\kappa_1,\kappa_2}(r)\equiv
 \sup_{({\cal A}, {\cal B})\in\, {\cal O}(r)} \sup_{\small
 \ba{c} A\in\sigma_{\kappa_1}({\cal A}),B\in\sigma_{\kappa_2}({\cal B})\\
\mu^B_0(B)>0\ea}
 \frac{|\mu^B_0(A\cap B)-\mu^B_0(A)\mu^B_0(B)|}{\mu^B_0(B)},\quad
\kappa_1,\kappa_2=0,1.
$$
%%-------------------------------------------------
%We say that the measure $\mu^B_0$ satisfies the strong uniform
% Ibragimov mixing condition if for any $d_1,d_2=0,1$,
 % $\varphi_{d_1,d_2}(r)\to 0$, $r\to\infty$,
%%--------------------------------------------
We assume that the measure $\mu^B_0$ satisfies the strong uniform
Ibragimov mixing condition, i.e., for any $\kappa_1,\kappa_2=0,1$,
  $\varphi_{\kappa_1,\kappa_2}(r)\to 0$, $r\to\infty$. Moreover,
$$
   \varphi_{\kappa_1,\kappa_2}(r)\le
C\nu_\kappa^2(r),\quad \mbox{where }
\,\,\kappa=\kappa_1+\kappa_2,\quad \kappa_1,\kappa_2=0,1.
  $$
%----------------------------------------
\begin{remark}\label{rQ}
{\rm (i) In \cite{DKS,DKM2}, we have constructed the generic
examples of the initial measures $\mu_0^B$ satisfying all
assumptions imposed.\\
 (ii) Condition {\bf S3} and the bound
(\ref{med'}) imply the bound (\ref{6.3'}).\\
 (iii) Condition (\ref{H2}) implies (\ref{med'}).
 Condition {\bf S3'} implies estimates (\ref{H2})
 with $i+|\alpha|\le 1$, $j+|\beta|\le 1$.
 The mixing condition {\bf S3'} is weaker than condition {\bf S3}.
  On the other hand, the estimates (\ref{H2}) with $\kappa>2$ are not
required for translation-invariant initial measures $\mu_0^B$ or
in the KGF case. \\
(iv) The conditions {\bf S2} and {\bf S3} admit various
modifications. We choose the variant which allows an application
to the case of the Gibbs measures $\mu_\pm^B$
 (see Section \ref{sec11.3} below).
%This is related to the slow long-range decay of the correlation
%function $Q_0^{00}(x,y)\sim|x-y|^{2-d}$, $|x-y|\to\infty$.
 }
\end{remark}

 %%%%%%%%%%%%%%%%%%%%%%%% 8.2  %%%%%%%%%%%%%%%%%%%%
\subsection{Convergence to equilibrium}
%%-----------------------------------------------------------

\begin{theorem} \label{l10.2}(see \cite{DKS, DKM2})
Let conditions {\bf A1}--{\bf A4} and all conditions imposed on $\mu_0^B$
in Section \ref{sec8.1} be satisfied.
 Then the assertions of Theorem \ref{l6.2} remains true with the
matrix $Q_\infty^B(x,y)=q^B_\infty(x-y)$ of the following form.
 In the Fourier transform,
  $ \hat q^B_{\infty}(k)
  =\hat q^+_{\infty}(k)+\hat q^-_{\infty}(k)$,
where (cf. (\ref{2.12'}))
$$\ba{rcl}
  \hat q^+_{\infty}(k)&=& \frac{1}{2}\Big(
\hat {\bf q}^+(k)+\hat C(k)\hat {\bf q}^+(k) \hat C^T(k)\Big),\\
 \hat q^-_{\infty}(k)&=& i\sgn(k_1)\frac{1}{2}\Big(\hat C(k)\hat {\bf q}^-(k)
-\hat {\bf q}^-(k)\hat C^T(k)\Big),   \ea
$$
  with ${\bf q}^+=(q_++q_-)/2$,
  ${\bf q}^-=(q_+-q_-)/2$, and $\hat C(k)$ from (\ref{C(k)}).
\end{theorem}
%---------------------------------
\begin{theorem}\label{t8.3}
Let conditions {\bf A1}--{\bf A5}, {\bf R1}--{\bf R3},
{\bf S0}, {\bf S1}, and all assumptions imposed on $\mu_0^B$ be satisfied.
Then the assertions of Theorem \ref{tA} hold.
%%% with $Q_\infty^B$ introduced in Theorem \ref{l10.2}.
\end{theorem}
%%-------------------------------------------------

This theorem can be proved in a similar way as Theorem \ref{tA}
(see Sections \ref{sec-com}--\ref{sec-char}).
\medskip

In Appendix A %%Section \ref{sec11.3}
 we will give an application of Theorems~\ref{l10.2} and \ref{t8.3} to the case when the measures $\mu_\pm^B$ from condition
 {\bf S2'} are Gibbs measures with different temperatures $T_+\not=T_-$.
 %%-------------------------- --------------------------
 \medskip\medskip

{\bf ACKNOWLEDGMENTS}

This work was supported partly by the research grant of RFBR
(Grant No. 12-01-00203).
The author is grateful to Alexander Komech for useful
discussions concerning several aspects of this paper.

\appendix
\setcounter{theorem}{0}
\setcounter{section}{1}
\setcounter{equation}{0}
%%%%%%%%%%%%%%%%%%%%%% 9  %%%%%%%%%%%%%%
\setcounter{equation}{0}
\section*{Appendix A: Gibbs measures}\label{sec9}
%%-------------------------

Here we study the case $L_B=\Delta-m^2$ only.
 Consider first
 the 'free' wave (or Klein--Gordon) equation,
\begin{equation}\label{freeKG}
    \left\{\ba{l}
\ddot \varphi_t(x) =(\Delta-m^2)\varphi_t(x),
 \quad t\in\R,\quad x\in\R^d,\\
\varphi_t(x)|_{t=0}=\varphi_0(x),\quad
\dot\varphi_t(x)|_{t=0}=\pi_0(x),
  \ea\right.
  \end{equation}
 where $m\ge0$, $d\ge 3$,  and $d$ is odd if $m=0$.
 Denoting $\phi_t=(\varphi_t,\pi_t)$, $t\in\R$,
  we rewrite (\ref{freeKG}) in the form
\begin{equation}\label{ap8.2}
 \dot \phi_t={\cal L}_B(\phi_t),\quad t\in\R, \quad
\phi_t|_{t=0}=\phi_0,
  \end{equation}
with ${\cal L}_B=\left( \begin{array}{cc} 0 & 1 \\
 \Delta-m^2 & 0 \end{array}\right)$.
%%---------------------------------------------------
In the Fourier transform representation, system (\ref{freeKG})
becomes
 $\dot{\hat\phi}_t(k)=\hat{\cal L}_B(k)\hat\phi_t(k)$,
 hence
  $\hat \phi_t(k)=\hat{\cal G}_t(k) \hat \phi_0(k)$,
  where
  $\hat{\cal G}_t(k)=\exp({\hat{\cal L}_B(k)t})$.
  Here we denote
$$
\hat{\cal L}_B(k)=
\left( \begin{array}{cc} 0 & 1 \\
 -\omega^2(k) & 0 \end{array}\right),\quad\quad
  \hat{\cal G}_t( k)=\left(
 \begin{array}{ccc}
  {\rm cos}~\omega t&\ds\frac{\sin \omega t}{\omega}  \\
 -\omega~{\rm sin}~\omega t &{\rm cos}~\omega t
 \end{array}\right),
  $$
   with $\omega\equiv\omega(k)=\sqrt{|k|^2+m^2}$.
Hence, the solution of (\ref{ap8.2}) is
 $ \phi_t=W^0 _t\phi_0={\cal G}_t(\cdot)*\phi_0$,
where
 ${\cal G}_t(x)=F^{-1}_{k\to x}[\hat{\cal G}_t(k)]$.
For simplicity of exposition, we omit below the index 0 in the notation
of the group $W^0_t$.

%%-----------------------------  9.1-----------------
\subsection{Phase space}
%%----------------------------------------------
We define the weighted Sobolev spaces
  with any $s,\alpha\in \R$.
%%----------------------------------------------------------

\begin{definition}
(i) $H^s_{\alpha}(\R^d)$ is the Hilbert space of the distributions
$\varphi\in S'(\R^d)$ with  finite norm
  \be\label{ws}
 \Vert \varphi\Vert_{s,\alpha}\equiv
  \Vert\langle x\rangle^{\alpha}
 \Lambda^s\varphi\Vert_{L^2(\R^d)}<\infty, \quad
\Lambda^s \varphi\equiv F^{-1}\left[\langle k\rangle^s
 \hat \varphi(k)\right],\quad s,\alpha\in\R.
  \ee
(ii) ${\cal H}^s_{\alpha} \equiv H_{\alpha}^{s+1}(\R^d)
 \oplus H_{\alpha}^s(\R^d)$  is the Hilbert space of pairs
   $\phi\equiv(\varphi(x),\pi(x))$  with  finite norm
 \be \label{1.5}
\Vert \phi\Vert_{s,\alpha}= \Vert\varphi\Vert_{s+1,\alpha}+
\Vert\pi\Vert_{s,\alpha}, \quad s,\alpha\in \R.
 \ee
(iii) ${\cal E}^s_{\alpha}\equiv{\cal H}^s_{\alpha}\oplus\R^d\oplus\R^d$
is the Hilbert space of vectors $Y\equiv(\phi(x),q,p)$ with finite norm
$$
  \Vert Y\Vert_{s,\alpha}=
\Vert\phi\Vert_{s,\alpha}+|q|+|p|, \quad s,\alpha\in\R.
  $$
\end{definition}
%----------------------------
Note that ${\cal H}^{\bar s}_{\bar\alpha}\subset{\cal H}^s_{\alpha}$
 (and also ${\cal E}^{\bar s}_{\bar\alpha}\subset{\cal E}^s_{\alpha}$) if
$\bar s>s$ and $\bar\alpha>\alpha$, and this embedding is compact.
 Moreover, for any $\alpha$, ${\cal H}^0_{\alpha}\subset{\cal H}$,
  ${\cal E}^0_{\alpha}\subset{\cal E}$
  (see Definition \ref{d1.1'}).

%%--------------------------------
\begin{lemma} Let $L_B=\Delta-m^2$,
$s,\alpha\in\R$, and conditions {\bf A1'} and {\bf A2} hold.
 Then (i) for every
$Y_0 \in {\cal E}^s_\alpha$, the Cauchy problem (\ref{1.1'})
has a unique solution $Y_t\in C(\R, {\cal E}^s_\alpha)$.\\
 (ii) For every $t\in \R$,
 the operator $S_t:Y_0\mapsto  Y_t$
 is continuous on ${\cal E}^s_\alpha$.
  Moreover,
  there exist positive constants $C_1,C_2>0$   such that
  $   \Vert S_t Y_0\Vert_{s,\alpha}
 \le C_1\langle t\rangle^{C_2}\Vert Y_0\Vert_{s,\alpha}$.
 \end{lemma}

This lemma can be proved by the similar technique from \cite{JP98}, where the nonlinear "wave field--particle" system was studied.

%%-------------------------  9.2  -------------------------
\subsection{Gibbs measures for the Klein-Gordon equation}
%%-------------------------------------------------------

Write $\phi=(\varphi,\pi)$. We introduce the (normalized) Gibbs
measures $g^{B}_\beta$
 on the space ${\cal H}^s_\alpha$.
 Formally,
 $$
  g^B_{\beta}(d\phi)=\frac{1}{Z_B}
  e^{-{\beta} H_B(\phi)}\prod_{x\in\R^d} d\phi(x),
  \quad H_B(\phi)=\frac12\int\left(|\nabla\varphi(x)|^2
   +m^2|\varphi(x)|^2+|\pi(x)|^2\right)\,dx.
 $$
  Now we adjust the definition of the Gibbs measures $g^B_{\beta}$.
Write $ \phi=(\phi^0,\phi^1)\equiv(\varphi,\pi), $ and denote by
$Q^{ij}(x,y)$, $i,j=0,1$, the correlation functions of
$g^B_{\beta}$,
$$
Q^{ij}(x,y)=\int\phi^i(x)\phi^j(y)\,g_\beta^B(d\phi)=
q^{ij}(x-y), \quad x,y\in\R^d.
$$
 We will define the
Gibbs measures $g^B_{\beta}$ as the Gaussian measures with the
correlation functions
 \be\label{q00-q11}
q^{00}(x-y)= T{\cal E}_m(x-y),~~ q^{11}(x-y)= T\delta(x-y),~~
q^{01}(x-y)= q^{10}(x-y)= 0,
 \ee
 where $T=1/\beta$,
 ${\cal E}_m(x)$ is the fundamental solution
  of the operator $-\Delta+m^2$.
 The correlation functions
$q^{ii}$ do not satisfy condition (\ref{med'}) because of
singularity at $x=y$. The singularity means that the measures
$g^B_{\beta}$ are not concentrated in the space $\cal H$.
%%---------------------------------------
\begin{definition}\label{def-bath}
For $\beta>0$, define the Gibbs measures $g^B_{\beta}(d\phi)$
 as the Borel probability measures
  $g^B_{\beta}(d\phi)=g^0_{\beta}(d\varphi)\times g^1_{\beta}(d\pi)$
  in ${\cal H}^s_{\alpha}=H^{s+1}_{\alpha}(\R^d)\otimes
  H^s_{\alpha}(\R^d)$, $s,\alpha<-d/2$, where $g^0_{\beta}(d\varphi)$
  and $g^1_{\beta}(d\pi)$ are  Gaussian Borel probability  measures
   in spaces $ H^{s+1}_{\alpha}(\R^d)$
and $H^s_{\alpha}(\R^d)$, respectively, with characteristic
functionals
 \beqn\label{10}
 \ba{c}\left. \ba{rcl}
  \hat g^0_{\beta}(f)&=&\ds\int
\exp\{i\langle \varphi,f\rangle\}\,g^0_{\beta}(d\varphi)=
\exp\left\{-\frac1{2\beta}\langle(-\Delta+m^2)^{-1}f,f\rangle
\right\}\\
 \hat g^1_{\beta}(f)&=& \ds\int
\exp\{i\langle\pi,f\rangle\}\,g^1_{\beta}(d\pi)
 =\exp\left\{-\frac1{2\beta}{\langle f,f\rangle}\right\}
\ea\right|\,\,\, f\in C_0^\infty(\R^d). \ea \eeqn
\end{definition}
%%--------------------------------------

By the Minlos theorem, the  Borel probability measures $g^0_\beta$
and $g^1_\beta$ exist in the spaces
  $ H^{s+1}_{\alpha}(\R^d)$ and $H^s_{\alpha}(\R^d)$, respectively,
 because {\it formally}
\be\label{Min}
  \int \Vert \varphi\Vert^2_{s+1,\alpha}\,
  g^0_{\beta}(d\varphi)<\infty, \quad
   \int\Vert \pi\Vert^2_{s,\alpha}\, g^1_{\beta}(d\pi)
  <\infty,\quad s,\alpha<-d/2.
  \ee
We verify (\ref{Min}). Definition (\ref{ws}) implies, for
  $\varphi\in{H}^{s}_{\alpha}(\R^d)$,
    \be\label{wsf}
     \Vert \varphi\Vert_{s,\alpha}^2=(2\pi)^{-2d}
\int_{\R^d} \langle x\rangle^{2\alpha} \Big(\int_{\R^{2d}}\ds
e^{-ix(k-k')}\langle k\rangle^{s}\langle k'\rangle^{s}
  \hat\varphi(k)\ov{\hat \varphi}(k')\,d k\, dk' \Big)dx.
  \ee
   Let $g(d\varphi)$ be a
 translation invariant measure in ${H}^{s}_{\alpha}(\R^d)$
with a correlation function $Q(x,y)=q(x-y)$. Let us introduce the
following correlation function
$$
   C(k,k')\equiv\int
\hat \varphi(k)\ov{\hat \varphi}(k')\, g(d\varphi)
  $$   in the sense of distributions. Since $\varphi(x)$
    is real-valued, we have
$$
C(k,k')=F_{x\to k}F_{x'\to-k'}Q(x,x')
 = (2\pi)^d\delta(k-k')\hat q(k).
$$
  Then, integrating (\ref{wsf}) with respect to the measure
$g(d\varphi)$, we obtain the formula
  $$
   \int \Vert\varphi\Vert_{s,\alpha}^2 \,g(d\varphi)
    =(2\pi)^{-d} \int_{\R^d} \langle x\rangle^{2\alpha}dx
  \int_{\R^d} \langle k\rangle^{2s}\hat q(k)\, dk.
   $$
      Substituting
 $\hat q(k)=T$ (see (\ref{q00-q11}))
  we obtain the second bound in (\ref{Min}).
 To obtain the first bound in (\ref{Min})
 we replace $s$ into $s+1$ and
put $\hat q(k)=T\hat {\cal E}_m(k)=T(|k|^{2}+m^2)^{-1}$.
\medskip

Below for spaces ${\cal E}^s_{\alpha}$ and
  ${\cal H}^s_{\alpha}$, we put $s,\alpha<-d/2$.
Definition \ref{def-bath}
implies the following lemma (cf the convergence (\ref{D.1})).
\begin{lemma}
The Gibbs measures $g_\beta^B$ are invariant w.r.t. $W_t$.
Moreover, the flow $W_t$ is mixing w.r.t. $g_\beta^B$.
\end{lemma}
%This lemma can be proved by the similar reasonings as in
%Section \ref{sec7}
%%------------------------------------------------------

In Section \ref{sec11.4},
we will define the Gibbs measures $g_\beta$ for the coupled system
and check the mixing property for the dynamics $S_t$ w.r.t. $g_\beta$.

%%------------------------  9.3 --------------------------
\subsection{Application of Theorems \ref{l10.2} and \ref{t8.3}
to Gibbs measures $\mu^B_\pm$}\label{sec11.3}
%----------------------------------

Let $\mu^B_{\pm}$ (see condition {\bf S2'}) be the Gibbs measures
$g^B_\pm\equiv  g^B_{\beta_\pm}$
 (with $\beta_\pm=1/T_\pm$) corresponding to different positive temperatures
$T_-\not= T_+$.
We define the Gibbs measures $g^B_{\pm}$ in
 the space ${\cal H}^s_{\alpha}$ (see Definition \ref{def-bath}) as the
Gaussian measures with the correlation functions (cf.
(\ref{q00-q11}))
 \be\label{4}
  q_{\pm}^{00}(x-y)=
T_{\pm}\,{\cal E}_m(x-y),~~ q_{\pm}^{11}(x-y)=
T_{\pm}\delta(x-y),~~ q_{\pm}^{01}(x-y)= q_{\pm}^{10}(x-y)= 0,
   \ee
 where $x,y\in\R^d$.
\medskip

Let us introduce $(\phi_-,\phi_+)$ as a unit random function in
the probability space $({\cal H}^s_\alpha\times {\cal H}^s_\alpha,
g^B_-\times g^B_+)$. Then $\phi_{\pm}$ are Gaussian independent
vectors in ${\cal H}^s_\alpha$. Define  a  Borel probability measure
$\mu^B_0\equiv g_0^B$ on ${\cal H}^s_\alpha$ as  the distribution of
the random function
$$
  \phi_0(x)=
\zeta_-(x_1)\phi_-(x)+\zeta_+(x_1)\phi_+(x),
 $$
  where functions $\zeta_\pm$ are introduced in (\ref{zeta}).
  Then correlation
functions of  $g^B_0$ are of the form (\ref{q1}) with
$q_{\pm}^{ij}$ from (\ref{4}).
Hence, the measure $g^B_0$  has zero mean and satisfies condition
(\ref{q1}) or (\ref{3}).
  However, $g^B_0$ does not satisfy (\ref{med'})
   because of singularity at $x=y$.
   Therefore, Theorem~\ref{l10.2} cannot be applied directly
to $\mu^B_0\equiv g^B_0$.
 The embedding ${\cal H}^s_{\alpha}\subset {\cal H}^s$
is continuous by the standard arguments of pseudodifferential
equations, \cite{H3}. The next lemma follows by Fourier transform
and the finite speed of propagation for the wave and Klein-Gordon
equation.
%%----------------------------------------------------
\begin{lemma}
The operators $W_t:\phi_0\mapsto \phi_t$ allow  a continuous
extension ${\cal H}^{s}\mapsto{\cal H}^{s}$.
\end{lemma}
%%%%%%%%%%%%%%%%%%%%%%%%%%%%%%%%%%%%%%%%%%%%%%

Let $\phi_0$ be the random function with the distribution $g^B_0$.
Hence $\phi_0\in {\cal H}^s_{\alpha}$ a.s.
 Denote by $g^B_t$ the distribution of $W_t\phi_0$.
 For the measures $g^B_t$, the following result was proved
 in \cite[Theorem~3.1]{DKS} and \cite[Section 4]{DKM2}.
%%------------------------------------------------
\begin{lemma}\label{t11.5}
Let $s<-d+{1}/{2}$. Then there exists a Gaussian  Borel
probability measure  $g^B_\infty$ on the space ${\cal H}^{s}$ such
that
  \be\label{1.8g}
 g^B_t\,{\buildrel{\hspace{2mm}{\cal H}^s}\over{-\hspace{-2mm}
\rightharpoondown}}\, g^B_\infty,\quad t\to \infty.\
 \ee
The correlation matrix $Q^B_{\infty}(x,y)=
 (q_{\infty}^{B,ij}(x-y))_{i,j= 0,1}$
 of the limit  measure $g^B_{\infty}$ has a form
\beqn
 \left. \ba{rcl}\ds
 q_{\infty}^{B,00}(x-y)&=&
  \frac{1}{2}(T_++T_-){\cal E}_m(x-y),\\
 q_{\infty}^{B,10}(x-y)&=&-q_{\infty}^{B,01}(x-y)=
\frac{1}{2}(T_+-T_-){\cal P}(x-y),\\
q_{\infty}^{B,11}(x-y)&=& \frac{1}{2}(T_++T_-)\delta(x-y),
\ea\right|\label{9.11}
\eeqn
  where ${\cal P}(x)=-iF^{-1}_{k\to x}
  \left[\sgn(k_1)/\omega(k)\right]$.
 In particular, the limiting mean
energy current density is formally
$$
\nabla q_\infty^{B,10}(0)=\frac{T_+-T_-}2\,\nabla{\cal P}(0)
 =-\frac{T_+-T_-}{2(2\pi)^d}
 \int_{\R^d}\frac{k\sgn(k_1)}{\sqrt{|k|^2+m^2}}\,dk=
 -\infty\cdot(T_+-T_-,0,\dots,0).
$$
The infinity means the 'ultraviolet divergence'.
\end{lemma}
%%---------------------------------------------------

Denote by $g_0$ a Borel probability measure on ${\cal E}^s_\alpha$
such that $P g_0=g_0^B$, where $P:(\phi_0,q_0,p_0)\in{\cal E}^s_\alpha\to\phi_0\in{\cal H}^s_\alpha$, and
 $g_0^B$ is the probability measure
on ${\cal H}^s_\alpha$ constructed above.
 Then, by the asymptotic behavior of $Y_t$ (see Section \ref{sec5})
and by Lemma~\ref{t11.5}, the following result holds
(cf. Theorems~\ref{tA} and \ref{t8.3}).
%%----------------------------------------------------
\begin{lemma}
Let $s<-d+{1}/{2}$.
Then the measures $g_t=S_t^*g_0$ weakly converge  to a limit measure
$g_\infty$ as $t\to\infty$ on the space ${\cal E}^{s}$.
The limit measure $g_\infty$ is Gaussian and its characteristic functional is
$\hat g_\infty(Z)=\exp\{-(1/2){\cal Q}_\infty(Z,Z)\}$,
where
${\cal Q}_\infty(Z,Z)=\langle q_\infty^B(x-y),\Pi(Z)\otimes\Pi(Z)\rangle$
with $q_\infty^B$ from (\ref{9.11}).
\end{lemma}
%%%----------------------------------------------------------

%%-------------------------9.4  ------------------------------
\subsection{Gibbs measure for the coupled system}\label{sec11.4}
%%--------------------------------------------------

 For $\beta>0$, we
introduce the (normalized) Gibbs measures $g_{\beta}$
 on the space ${\cal E}^s_\alpha$.
 Formally,
 $$
 g_{\beta}(d\phi\, d\xi)=\frac{1}{Z} e^{-\beta
H(\phi,\xi)}\prod\limits_{x\in\R^d} d\phi(x)\,d\xi.
 $$
%%--------------------------------
\begin{definition}\label{d11.8}
 For $\beta>0$, define the Gibbs measures
 $g_{\beta}(d\phi\, d\xi)$ in ${\cal E}^s_\alpha$, $s,\alpha<-d/2$,
 as
  \beqn \label{G}
g_{\beta}(d\phi\, d\xi)= \frac{1}{Z} e^{-\beta q\cdot\langle\rho,
\nabla\varphi\rangle}\, g^{B}_{\beta}(d\phi)\times
g^{A}_{\beta}(d\xi).
 \eeqn
  Here $\beta=1/T$ is an inverse temperature,
  $g^{B}_{\beta}(d\phi)$ is defined in Definition \ref{def-bath},
  and %$Z$ is the normalization constant,
 $g^{A}_{\beta}$  is the  Gibbs   measure on $\R^d\times\R^d$,
 \be\label{1.10'}
  g^{A}_{\beta}( d\xi)  =
 \frac{1}{Z_A} e^{-\beta H_A(\xi)}\, d\xi,\quad
 H_A(\xi)=\frac12(|p|^2+\omega^2 |q|^2).
\ee
\end{definition}
%%%--------------------------------------------

In Section \ref{sec11.5} we will prove the invariance
 of the Gibbs measures $g_\beta$ w.r.t. the group $S_t$.
%%%-------------------------------------------------------------
\begin{lemma}
The flow $S_t$ is mixing w.r.t. $g_\beta$, i.e.,
 for any functions
$F_1,F_2\in L_2({\cal E}^s_\alpha,g_\beta)$, we have
 $$
\int F_1(S_tY)F_2(Y)g_\beta(dY)\to
 \int F_1(Y)g_\beta(dY)\int F_2(Y)g_\beta(dY)
 \quad \mbox{as }\,\,t\to\infty.
 $$
\end{lemma}
%%------------------------------------------
{\bf Proof}\, It suffices to check that
 for any $Z_1,Z_2\in{\cal D}$,
 \be\label{11.12}
\int\langle S_tY_0,Z_1\rangle\langle Y_0,Z_2\rangle
\,g_\beta(dY_0) \to0,\quad t\to\infty.
 \ee
Let $Z_1=(f,u,v)\in{\cal D}$. By Corollary \ref{cor3.2} and
formulas (\ref{al})--(\ref{teta}), we obtain
$$ |q_t-\langle W_t\phi_0,\alpha\rangle|
   +|p_t-\langle W_t\phi_0,\beta\rangle|  \le
     C_1\ve_m(t)|\xi_0|+
C_2\sqrt{\tilde\ve_m(t)}\sup_{\tau\in\R} |\langle
W_\tau\phi_0,\nabla\rho_0\rangle|,
   $$
   and
$$
   |\langle \phi_t,f\rangle-\langle W_t\phi_0,f_*\rangle|\le
   C_1\ve_m(t)|\xi_0|+C_2\sqrt{\tilde\ve_m(t)}
    \left(\sup_{\tau\in\R} |\langle
W_\tau\phi_0,\nabla\rho_0\rangle|+\sup_{\tau\in\R} |\langle
W_\tau\phi_0,\alpha\rangle|\right).
 $$
 These bounds can be proved similarly to Proposition \ref{l7.1}.
 Hence, to prove (\ref{11.12}) it suffices to verify that
\be \label{11.13}
  \int\langle\phi_0, W'_t\chi\rangle\langle Y_0,
Z_2\rangle \,g_\beta(dY_0)\to0 \quad\mbox{as }\,\,\,t\to\infty,
 \ee
 with $\chi=\alpha,\beta, f_*$.
Since
  \be\label{6.5}
  F_{x\to k}[W'_tf]=\left(\ba{cc}
  \cos\omega(k)t&-\omega(k)\sin\omega(k)t\\
  \omega^{-1}(k)\sin\omega(k)t&\cos\omega(k)t
  \ea\right)\left(\ba{c}\hat f^0(k)\\\hat f^1(k)
  \ea\right),
  \ee
 then Definition {\ref{d11.8}}, equalities (\ref{q00-q11}),
     and the Lebesgue--Riemann theorem imply (\ref{11.13}).
 \bo

%%----------------------------------
\subsection{Effective Hamiltonian}
%---------------------------------
To prove the invariance of the Gibbs measures $g_\beta$
we use notations introduced by Jak\v{s}i\'c and Pillet in \cite{JP98}.
At first, we rewrite the system (\ref{cs1})--(\ref{cs2}) in new variables.
Introduce an {\em effective potential} by
\be\label{Veff}
V_{eff}(q)=
 \frac12(\omega^2|q|^2-q\cdot K_m q),
\ee
where $K_m$ is the 'coupling constant matrix' defined in (\ref{Km}).
 By condition {\bf R1'}, $V_{eff}(q)\ge0$.
 Define  $\R^d$-valued function $h(x)$,
\be\label{h}
h(x)=(\Delta-m^2)^{-1}\nabla\rho(x),  %=-{\cal E}*\nabla\rho(x),
\quad x\in\R^d,
\ee
  where $\rho$ is the coupled function, and put
$h_0=(h,0)\in\R^d\times\R^d$. Then  the fist equations in
  (\ref{1''}) become
  \be\label{9.9}
 \dot\phi_t={\cal L}_B\phi_t+q_t\cdot\nabla\rho^0=
 {\cal L}_B(\phi_t+q_t\cdot h_0),\quad\mbox{with }\,\,
  {\cal L}_B=\left(\ba{cc}0&1\\\Delta-m^2&0\ea\right),
    \ee
 because ${\cal L}_B h_0=(0,\nabla\rho)$.
 Define a $\R^2$-valued function
  $\psi\equiv\psi(x)$, $x\in\R^d$,
   where
$$
 \psi=(\psi^0,\psi^1):
 \quad \psi^0=\varphi+ q\cdot h,\quad \psi^1=\pi.
 $$
Then (\ref{9.9}) becomes
$\dot\psi_t={\cal L}_B\psi_t+\dot q_t\cdot h_0$.
 Recall that ${\cal L}_B$ is the generator of the
group $W_t$. Hence, in new variables $(\psi_t,\xi_t)$ the system
(\ref{cs1})--(\ref{cs2})  becomes
 \beqn
 \psi_t&=&W_t\psi_0+\int\limits_0^tW_{t-s}h_0\cdot\dot q_s\,ds,
 \quad x\in\R^d,\quad t\in\R,\nonumber\\
\ddot q_t&=&-\nabla V_{eff}(q_t)-\int\limits_0^t\Gamma(t-s)\dot
q_s\,ds+{\cal F}(t),\label{Langevin}
 \eeqn
where
 $ {\cal F}(t):=\langle\nabla\rho_0,W_t\psi_0\rangle$,
  $\nabla V_{eff}(q_t)=(\omega^2I-K_m)q_t$,
   the matrix $K_m$ is defined in (\ref{Km}), its entries are
$$
K_{m,ij}=-\langle\nabla_i\rho_0,h_0^j\rangle= (2\pi)^{-d}\int
\frac{k_ik_j|\hat\rho(k)|^2}{k^2+m^2}\,dk,
$$
and $\Gamma(t)$ stands for the $\R^d\times\R^d$ matrix with
entries $\Gamma_{ij}(t)$,
 \be\label{gamma}
\Gamma_{ij}(t):=-\langle\nabla_i\rho_0,W_t h_0^j\rangle
 =(2\pi)^{-d}\int
 k_ik_j\frac{\cos\omega(k)t}{\omega^2(k)}\,|\hat\rho(k)|^2\,dk,
 \quad i,j=1,\dots,d.
 \ee
The equation (\ref{Langevin}) is called the {\it generalized or
retarded Langevin equation} with the {\it random force} ${\cal
F}(t)$ and with the {\it memory kernel} $\Gamma(t)$.
%%-----------------------------------------------------
\begin{remark}
{\rm
(i) By (\ref{gamma}), we have $\Gamma(0)=K_m$. Moreover,
$\dot\Gamma_{ij}(t)=-D_{ij}(t)$, where $D_{ij}(t)$ are the entries
of the matrix $D(t)$ defined in (\ref{3.4}).\\
(ii)  $\rho_0(x)=h_0(x)=0$ for
$|x|\ge R_\rho$, by condition {\bf R2}.
Hence, $\Gamma(t)=0$ for $|t|>2R_\rho$ if $m=0$
  due to a {\it strong} Huyghen's principle,
and $|\Gamma(t)|\le C(1+|t|)^{-d/2}$ if $m\not=0$.\\
 (iii)
 It follows from (\ref{q00-q11}), (\ref{6.5})  and (\ref{gamma}) that
 $\int\left({\cal F}(t)\otimes{\cal F}(s)\right)g_\beta^B(d\psi)=
(1/\beta)\,\Gamma(t-s)$ ({\it fluctuation--dissipation} relation).\\
 % This is an example of the fluctuation-dissipation theorem.
 (iv) The force ${\cal F}(t)$ equals $F(t)-\Gamma(t)q_0$ with $F(t)$
from (\ref{cs2}).
}\end{remark}
%%-------------------

 Introduce an {\em effective Hamiltonian}
$ H_{A}^{eff}(\xi)=|p|^2/2+V_{eff}(q)$.
 Hence, by (\ref{Ham}),
$$
H(\phi,\xi)=H_B(\psi)+H_{A}^{eff}(\xi).
$$
%%%%%%%%%%%%%%%%%%%%%%%%%%%%%%%%%%%%%
\begin{definition}\label{dir}
(i) Define a map ${\bf T}$ on ${\cal E}^s_\alpha$ by the rule
 $$
  {\bf T}:(\phi,\xi)\to(\psi,\xi),\quad \psi=\phi+q\cdot h_0.
$$
   (ii)
Denote $g^{\bf T}_\beta(d\psi d\xi):=g_\beta({\bf T}^{-1}(d\psi
d\xi))$. Then, $g_\beta^{\bf T}(d\psi d\xi)
=g_\beta^B(d\psi)\times g_\beta^{eff}(d\xi)$, where
 $g_\beta^B$ is defined in Definition~\ref{def-bath},
 $g_\beta^{eff}$ is a Gaussian measure defined by
$g_\beta^{eff}(d\xi)=(1/Z)e^{-\beta H_A^{eff}(\xi)}d\xi$.
\end{definition}

%%%-------------------------------------------------------
\subsection{Invariance of Gibbs measures $g_\beta$}\label{sec11.5}
%%%----------------------------------------------

\begin{pro}
Let conditions {\bf R1} and {\bf R2} hold. Then the Gibbs measures
$g_{\beta}$, $\beta>0$,
 are invariant with respect to the dynamics, i.e.
 \be\label{1.15}
S^*_t g_{\beta}(\omega):=
g_{\beta}(S_t^{-1}\omega)=g_{\beta}(\omega),\quad
 \mbox{for }\,
 \omega\in {\cal B}({\cal E}^s_\alpha)\quad \mbox{and }\,\, t\in\R.
  \ee
 Here ${\cal B}({\cal E}^s_\alpha)$ is the Borelian $\sigma$-algebra of subsets in ${\cal E}^s_\alpha$.
\end{pro}
%%------------------------------------------------
 %Note that the invariance (\re{1.15}) of Gibbs measures
 %is natural from  a formal definition (\ref{G}).
%%---------------------------------------------------------------------
{\bf Proof}\, For simplicity,  we omit indices $\alpha,s$
 in notations ${\cal E}^s_{\alpha}$ and ${\cal H}^s_{\alpha}$.
%  The maps $S_t$, $t\in\R$, are continuous isomorphisms of ${\cal E}$, hence
The invariance (\ref{1.15}) is equivalent to the identity:
 \be\label{10.1}
  \frac d{dt}\int_{\cal E}F(S_t Y) g_\beta(dY)=0,
   \quad t\in\R,
 \ee
   for any bounded continuous functional $F(Y)$ on ${\cal E}$, i.e., $F(Y)\in C_b({\cal E})$.
   It suffices to prove (\ref{10.1})
  with $t=0$ only. Indeed, since $S_{t+\tau}=S_t~S_\tau$,
we have
 \be\label{10.4}
\frac d{d\tau}\int_{\cal E}F(S_\tau S_t Y) g_\beta(dY)=
 \frac d{d\tau}\int_{\cal E}F(S_{\tau+t} Y) g_\beta(dY)=
\frac d{dt}\int_{\cal E}F(S_{t} S_\tau Y) g_\beta(dY).
 \ee
Let $\frac d{d\tau}\int_{\cal E}F(S_\tau Y) g_\beta(dY)\Big|_{\tau=0} =0$.
Since $F(S_tY)\in C_b({\cal E})$, then
$\frac d{d\tau}\int_{\cal E}F(S_\tau S_t Y) g_\beta(dY)\Big|_{\tau=0}=0$
 for any fixed $t\in\R$. Hence, (\ref{10.4}) implies
$$
0=\frac d{d\tau}\int_{\cal E}F(S_\tau S_t Y) g_\beta(dY)\Big|_{\tau=0}
%= \frac d{dt}\int_{\cal E}F(S_{t} S_\tau Y) g_\beta(dY)\Big|_{\tau=0}
 =\frac d{dt}\int_{\cal E}F(S_{t} Y) g_\beta(dY),
$$
and (\ref{10.1}) follows.
Moreover, it suffices to verify (\ref{10.1}) with $t=0$ and
$F(Y)=\exp(i\langle Y,Z\rangle)$ for every
$Z=(f_0(x),f_1(x),u,v)\in {\cal D}$.
  Then, by (\ref{1.1'}), identity (\ref{10.1}) with $t=0$ becomes
 \be\label{10.2}
\frac d{dt}\int_{\cal E}e^{i\langle S_tY,Z\rangle}
g_\beta(dY)\Big|_{t=0}=
 \int_{\cal E}e^{i\langle Y,Z\rangle}i
\langle {\cal L}(Y),Z\rangle  g_\beta(dY)=0,
 \ee
where
  \be\label{9.17}
 {\cal L}(\varphi,\pi,q,p)
 =(\pi,(\Delta-m^2)\varphi+q\cdot\nabla\rho,p,-\omega^2q
  +\langle \nabla\rho,\varphi\rangle).
 \ee
 Now we prove (\ref{10.2}). Denote by $I$ the integral
$$
I:=\int_{\cal E}e^{i\langle Y,Z\rangle}i \langle {\cal
L}(Y),Z\rangle  g_\beta(dY),
$$
and check that $I=0$.
 Definition~\ref{dir} implies
$ g_\beta(dY)=g^{\bf T}_\beta({\bf T}dY)$. Hence,
 \be\label{10.3}
\int\limits_{\cal E}F(Y)\, g_\beta(dY)
 %%=\int\limits_{\cal E}F\left({\bf T}^{-1}(\psi,\xi)\right)\,
 %%g^{\bf T}_\beta\left(d\psi d\xi\right)
 =\int\limits_{\R^{2d}} g^{eff}_\beta(d\xi)
 \int\limits_{{\cal H}}
F(\psi- q\cdot h_0,\xi)\, g^B_\beta(d\psi).
 \ee
 Using (\ref{9.17}), (\ref{10.3}), and (\ref{h}),
   we rewrite $I$ in the form
 \beqn\nonumber
 \ba{rcl}
I&=&\ds\int\limits_{\R^{2d}}
   e^{i(u\cdot q+v\cdot p)} g^{eff}_\beta(d\xi)
  \int\limits_{{\cal H}} e^{i\langle\psi^0-
q\cdot h,f_0\rangle+i\langle\psi^1,f_1\rangle} \Big(
i\langle\psi^1,f_0\rangle
 +i\langle(\Delta-m^2)\psi^0,f_1\rangle\\
&&+iu\cdot p+iv\cdot[-\omega^2q+\langle\nabla\rho,\psi^0-q\cdot
h\rangle] \Big)g^0_\beta(d\psi^0) g^1_\beta(d\psi^1).
 \ea \eeqn
  Integrals over Gaussian measures
$g^0_\beta(d\psi^0)$ and $g^1_\beta(d\psi^1)$ can be represented
as variational derivatives of their characteristic functionals
$\hat g^0_\beta(f_0)$ and $\hat g^1_\beta(f_1)$:
$$
\int e^{i\langle\psi,f\rangle}i\langle \psi,\cdot\rangle g^i_\beta
(d\psi)=\langle \frac{\delta}{\delta f}\,\hat
g^i_\beta(f),\cdot\rangle, \quad i=0,1,\quad f\in
C_0^\infty(\R^d).
$$
Then
 \beqn \label{10.6} \ba{rcl}
  I&=&\ds\int\limits_{\R^{2d}} e^{i(u\cdot q+v\cdot p)}
e^{-i q\cdot\langle h,f_0\rangle} \Big(\langle\frac\delta {\delta
f_1},f_0\rangle+ \langle(\Delta-m^2) \frac\delta {\delta f_0}, f_1\rangle\\
&&\ds+iu\cdot p+v\cdot\Big[-i\omega^2 q+
 \langle\frac\delta {\delta f_0},\nabla\rho\rangle
 -i \langle \nabla\rho,q\cdot h\rangle  \Big]\Big)
 \hat g^0_\beta(f_0) \hat g^1_\beta(f_1)
  \,g^{eff}_\beta(d\xi).
 \ea\eeqn
Using (\ref{10}), we calculate
 \beqn\nonumber
\left. \ba{ll}
 \ds\langle\frac\delta {\delta f}\,\hat g^0_{\beta}(f),\cdot\rangle=
-\frac 1{\beta}\,e^{-\frac{1}{2\beta}\langle (-\Delta+m^2)^{-1}f,
f\rangle}\langle (-\Delta+m^2)^{-1} f,\cdot\rangle\\
\ds\langle\frac\delta {\delta f}\,\hat g^1_{\beta}(f),\cdot\rangle
 = -\frac1{\beta}\,
  e^{-\frac{1}{2\beta}\langle f,f\rangle}\langle f,\cdot\rangle
 \ea
\right|\quad f\in C_0^\infty(\R^d).
 \eeqn
  Therefore, we reduce (\ref{10.6}) to the following integral
 \beqn
  I&=&C \int\limits_{\R^{2d}}
e^{i(u\cdot q+v\cdot p)} e^{-i q\cdot\langle h,f_0\rangle}
 \Big(
iu\cdot p-iv\cdot\nabla V_{eff}(q)+\frac{1}\beta\, v\cdot
\langle f_0,h\rangle \Big)e^{-\beta H_A^{eff}(\xi)}\,d\xi\nonumber\\
&=& C_1\int\limits_{\R^{2d}}
  e^{i(u\cdot q+v\cdot p)} (u\cdot\nabla_p-v\cdot\nabla_q)
 \Big[e^{-iq\cdot\langle h,f_0\rangle-\beta H_A^{eff}(q,p)}
 \Big]\,dq\,dp,\nonumber
 \eeqn
 by (\ref{Veff}) and (\ref{h}).
 Partial integration in $q$ and  in $p$ leads to
$$
I=C_2 \int\limits_{\R^{2d}} e^{i(u\cdot q+v\cdot p)}(-u\cdot(iv)+
v\cdot(iu)) e^{-i q\cdot\langle h,f_0\rangle-\beta
H_A^{eff}(q,p)} \,dqdp=0.\,\,\,\,\bo
$$

\appendix
\setcounter{theorem}{0}
\setcounter{section}{2}
\setcounter{equation}{0}
%%%%%%%%%%%%%%%%%%%%%% 9  %%%%%%%%%%%%%%
\setcounter{equation}{0}
\section*{Appendix B: Existence of solutions}
%%-------------------------

 Proposition \ref{p1.1'} can be proved by using the
methods of  \cite[Lemma 6.3]{KSK}.
In this section, we outline the proof of this proposition.\\
%%-----------------------------------------
% Now we prove the axillary lemma.
%%--------------------------------------------
%\begin{lemma}\label{l3.1}
%Let conditions {\bf A1}--{\bf A4}, {\bf R1} and {\bf R2} hold. Then\\
%(i) for every $Y_0\in E$, the Cauchy problem (\ref{1.1'})
%has a unique solution $Y_t\in C(\R,E)$.\\
%(ii) For every $t\in \R$, the operator $S_t:Y_0\mapsto Y_t$
%is continuous on $E$.\\
%(iii) The energy is conserved and finite,
%\be\label{3.0}
%H(Y_t)=H(Y_0)\,\,\,\mbox{for }\,\,t\in\R.
%  \ee
%\end{lemma}
%%------------------------------------------------
{\bf Proof of Lemma \ref{l3.1}.}
{\it Step (i)}
 If $\rho=0$, then the
existence and uniqueness of the solution $Y_t\in C(\R,E)$ to
problem (\ref{1.1'}) is well-known
(see, for example, \cite{Mikh}).  % [Lions, Madgenes]
Represent the solution $Y_t$ as the pair of the functions
$(Y^0_t,Y^1_t)$, where $Y^0_t=(\varphi_t,q_t)$,
$Y^1_t=(\pi_t,p_t)$.
Therefore, problem (\ref{1.1'}) for $Y_t\in C(\R,E)$ is equivalent
to
\be\label{intD}
  Y_t=e^{{\cal L}_0 t} Y_0
  +\int\limits_0^t e^{{\cal L}_0(t-s)}BY_s\,ds,
  \ee
 where $Y_0=(\varphi_0,q_0,\pi_0,p_0)\in
 E=H^1_F(\R^3)\otimes \R^3\otimes L^2(\R^3)\otimes \R^3$,
   \beqn\label{A_0}
  \left.\ba{cc} {\cal L}_0=\left( \ba{cc}0&I\\
  {\cal A}_0&0 \ea\right),
 & {\cal A}_0\left(\ba{c}\varphi\\q\ea\right)
  =\left(\ba{c}  L_B \varphi\\-\omega^2 q\ea\right)\\
 B(Y^0,Y^1)=(0,RY^0),&
RY^0:=\Big(q\cdot\nabla\rho,
 \langle\varphi,\nabla\rho\rangle\Big)
 \ea\right|\quad\mbox{for }\,\, Y^0=(\varphi,q),
  \eeqn
(cf (\ref{7.4})). Note that
  $\Vert e^{{\cal L}_0t}Y_0\Vert_E\le C\Vert Y_0\Vert_E$;
and the second term in (\ref{intD}) is estimated by
$$
\sup\limits_{|t|\le T}\Vert \int\limits_0^t e^{{\cal L}_0 (t-s)}
BY_s\,ds\Vert_E
   \le C\,T\sup\limits_{|s|\le T}\Vert Y_s\Vert_E.
$$
This bound and the contraction mapping principle imply the
existence and uniqueness of the local solution $Y_t\in
C([-\ve,\ve],E)$ for some $\ve>0$.
\medskip

{\it Step (ii)} To prove the energy conservation
\be\label{3.0}
H(Y_t)=H(Y_0)\,\,\,\mbox{for }\,\,t\in\R,
  \ee
 we first assume that
$\phi_0=(\varphi_0,\pi_0)\in C^3(\R^3)\times C^2(\R^3)$ and
$\phi_0(x)=0$ for $|x|\ge R_0$. Then $\varphi_t(x)\in
C^2(\R^3_x\times\R_t)$ and
$$
\varphi_t(x)=0\quad\mbox{for }\,
  |x|\ge |t|+\max\{R_0,R_a,R_\rho\}
$$
by the integral representation (\ref{intD}) and
  conditions {\bf A2} and {\bf R2}.
   Therefore, for such initial data, relation (\ref{3.0})
can be proved by integrating by parts. Hence, for $Y_0\in E$,
(\ref{3.0}) follows from the continuity of $S_t$
  and from the fact that
 $C_0^3(\R^3)\oplus\R^3\oplus C_0^2(\R^3)\oplus\R^3$
is dense in $E$.
\medskip

{\it Step (iii)} In the case of WF, we apply condition {\bf A3} and obtain
 \beqn
&&
\frac12\int\Big(
\sum\limits_{ij}\nabla_i\varphi(x)a_{ij}(x)\nabla_j\varphi(x)+
2 q\cdot\nabla\varphi(x)\rho(x)\Big)\,dx \nonumber\\
&&\ge
\frac12\int \Big(\alpha|\nabla\varphi(x)|^2+
2 q\cdot\nabla\varphi(x)\rho(x)\Big)\,dx
=\frac{\alpha}{2} \Vert \nabla\varphi+\frac{q\rho}{\alpha}\Vert^2
-\frac1{2\alpha}{|q|^2}\Vert\rho\Vert^2,
 \nonumber
 \eeqn
where $\Vert\cdot\Vert$ stands for the norm in $L^2$.
 In the case of KGF,
$$
 \frac12\int\Big( m^2|\varphi(x)|^2-2 \varphi(x)q\cdot\nabla\rho(x)\Big)\,dx
\ge \frac12\, m^2\Vert\varphi-\frac{q\cdot\nabla\rho}{m^2} \Vert^2-
\frac1{2m^2}|q|^2 \Vert\nabla\rho\Vert^2.
 $$
  Hence, the Hamiltonian functional $H(Y)$ is nonnegative. Indeed,
in the case of WF,
\beqn\label{positive}
 H(Y)&\ge&\frac12\int\Big(
|\pi(x)|^2+\alpha\Big|\nabla\varphi(x)+\frac{q\rho(x)}{\alpha}\Big|^2
+a_0(x)|\varphi(x)|^2\Big)\,dx\nonumber\\
&&+\frac12\Big(\omega^2-\frac1\alpha\Vert\rho\Vert^2\Big)|q|^2
  +\frac12|p|^2\ge0
\eeqn
 by condition {\bf R1}.
In the case of KGF,
\beqn\label{positive1}
 H(Y)&\ge&\frac12\int\Big(|\pi(x)|^2
 +\sum\limits_j\left|(\nabla_j-i A_j(x))\varphi(x)\right|^2+
  m^2\Big|\varphi(x)-\frac{q\cdot\nabla\rho(x)}{m^2}\Big|^2 \Big)\,dx\nonumber\\
&&+\frac12\Big(\omega^2-\frac1{m^2}\Vert\nabla\rho\Vert^2\Big)|q|^2
  +\frac12|p|^2\ge0
\eeqn
by condition {\bf R1}.
Moreover, by (\ref{3.0}), %for $|t|<\ve$ and
 (\ref{positive}) and (\ref{positive1}),
  we obtain
 \be\label{B.6}
 \Vert Y_t\Vert^2_{E}\le C\, H(Y_t)=C\, H(Y_0).
  \ee
 On the other hand, in the case of KGF, we have
 \beqn\label{B.7}
H(Y)&\le&\frac12\Big\{\sum_j\Vert (\nabla_j-iA_j(x))\varphi\Vert^2+
\Vert\nabla\varphi\Vert^2
  + \Vert\pi\Vert^2 +m^2\Vert\varphi\Vert^2+
(\omega^2+\Vert\rho\Vert^2)
  |q|^2+|p|^2\Big\}\nonumber\\
  &\le& C\Vert Y\Vert^2_{E},
\eeqn
since
  $|q\cdot\langle\nabla\varphi,\rho\rangle|\le
(\Vert\nabla\varphi\Vert^2+|q|^2\Vert\rho\Vert^2)/2$.
In the WF case,
$$
H(Y)\le C\left(\Vert\nabla\varphi\Vert^2  + \Vert\pi\Vert^2 +
(\omega^2+\Vert\rho\Vert^2)
  |q|^2+|p|^2+\int a_0(x)|\varphi(x)|^2\,dx\right).
$$
Since $Y\in E$,  $\varphi\in H^1_F$.
For the WF case, $H^1_F$ is the completion of real space
$C^\infty_0(\R^3)$ with the norm $\Vert\nabla\varphi\Vert$.
Therefore,
$H^1_F=\{\varphi\in L^6(\R^3):\,|\nabla\varphi|\in L^2\}$
by  Sobolev's embedding theorem.
Hence,
$$
\int a_0(x)|\varphi(x)|^2\,dx\le
 C \Vert \varphi\Vert_{L^6}^2\le C_1\Vert\nabla\varphi\Vert^2.
$$
Using (\ref{B.6}) and (\ref{B.7}),
we obtain the {\it a priori} estimate
   \beqn\label{enest}
  \Vert Y_t\Vert_{E}\le
C_1\Vert Y_0\Vert_E\quad\mbox{for }\,\, t\in\R.
  \eeqn
   Therefore, properties (i)--(iii) of Lemma \ref{l3.1}
 for arbitrary $t\in\R$
follow from bound (\ref{enest}). \bo
\medskip

%---------------------------
We return to the proof of Proposition \ref{p1.1'}. Let us choose
$R>\max\{R_a,R_\rho\}$ with $R_a$ and $R_\rho$
from conditions {\bf A2} and {\bf R2}. Then, by the
integral representation (\ref{intD}), the solution $Y_t$ for
$|x|<R$ depends only on the initial data $Y_0(x)$ with
$|x|<R+|t|$.
   Thus, the continuity of $S_t$ in ${\cal E}$ follows
from the continuity in $E$.

For every $R>0$, define  the local energy seminorms by
$$%\la{3.4'}
  \Vert Y\Vert^2_{E(R)}:=
 \int\limits_{|x|<R} \Big(|\nabla
\varphi(x)|^2+m^2|\varphi(x)|^2+|\pi(x)|^2\Big)\,dx
  +|q|^2+|p|^2, \quad Y=(\varphi,\pi,q,p),
$$
 where $m>0$ for the KGF case, and $m=0$ for the WF case.
   By estimate (\ref{enest}), we
obtain the following local energy estimates:
 $$
 \Vert S_t Y_0\Vert^2_{E(R)}\le C\Vert Y_0\Vert^2_{E(R+|t|)}
\quad \mbox{for }\,R>\max\{R_\rho,R_a\}\quad \mbox{and }\,\,t\in\R.
 $$
  Hence,
 for any $T>0$ and $R>\max\{R_\rho,R_a\}$,
$$
  \sup\limits_{|t|\le T}\Vert S_t Y_0\Vert_{{\cal E},R}
\le C(T)\Vert Y_0\Vert_{{\cal E},R+T}.\quad\bo
$$

\newpage
%%%%%%%%%%%%%%%%%%%%%%%%%%%%%%%%%%%%%%%%%%%%%%%%%%%%%%%%%
%%%%%%%%%%                  bibliography
%%%%%%%%%%%%%%%%%%%%%%%%%%%%%%%%%%%%%%%%%%%%%%%%%%%%%%

\end{document}